\documentclass[aps,prb,twocolumn]{revtex4} 

\usepackage{graphicx}
\usepackage{dcolumn}
\usepackage{bm}

\newcommand{\comment}[1]{}

\begin{document}
\renewcommand{\theequation}{\arabic{section}.\arabic{equation}}

\title{Bose-Einstein Condensation, the Lambda Transition,
and Superfluidity for Interacting Bosons}


\author{Phil Attard}
\affiliation{{\tt phil.attard1@gmail.com}}


\begin{abstract}
Bose-Einstein condensation and the $\lambda$-transition
are described in molecular detail for bosons interacting
with a pair potential.
New phenomena are identified
that are absent in the usual ideal gas treatment.
Monte Carlo simulations of Lennard-Jones helium-4
neglecting ground momentum state bosons
give a diverging heat capacity approaching the transition.
Pure permutation loops give continuous growth
in the occupancy of the ground momentum state.
Mixed ground and excited momentum state permutation loops
give a discontinuous transition to the condensed phase.
The consequent latent heat for the $\lambda$-transition
is 3\% of the total energy.
The predicted critical velocity for superfluid flow
is within a factor of three of the measured values
over three orders of magnitude in pore diameter.
\end{abstract}


\maketitle

%
\section{Introduction}
\setcounter{equation}{0} \setcounter{subsubsection}{0}
%

The spike in the heat capacity and the onset of superfluidity
in liquid helium-4 at 2.17\,K is known as the $\lambda$-transition.
London (1938) suggested that it was due to
Bose-Einstein condensation,
based upon analysis of the ideal gas
that also showed a peak in the heat capacity
as the ground energy state first begins to fill at 3.1\,K.
Using this ideal gas model
Tisza (1938) developed a two-fluid
model for superfluidity in which the ground energy state bosons
were said to flow without viscosity,
transparent to the excited  energy state bosons
that were said to have the ordinary viscosity.
Landau (1941) modified Tisza's (1938) model
to include collective behavior for interacting bosons
as quantum excitations (phonons).
He postulated the existence of rotons,
whose energy spectrum he fitted to measured heat capacity data.
Later theoretical work focussed on interpreting rotons as quantized vortices
(Feynman 1954,
Kawatra and Pathria 1966, Pathria 1972),
although direct evidence for this is lacking.
Accounts of the fascinating history of the $\lambda$-transition
and superfluidity have been given by Donnelly (1995, 2009)
and by Balibar (2014, 2017).

The main limitation of London's (1938) and Tisza's (1938) approaches
is that they use the ideal gas,
which neglects the interactions between the helium atoms,
which very interactions are responsible for the liquid state.
Landau's (1941) attempt to include interactions via phonons and rotons
may be criticized as lacking a specific molecular basis,
and as being disconnected from any phase transition,
including Bose-Einstein condensation.
In addition all three approaches lack a derivation of superfluidity
that explains the absence of viscosity
in the presence of molecular interactions (i.e.\ collisions).

The present paper analyzes the $\lambda$-transition
for interacting bosons.
Section~\ref{Sec:Ideal}
gives a re-analysis of London's (1938) ideal gas approach
with the focus on a particular approximation
that separates the ground energy state from the continuum integral
for the excited states.
Here and in the rest of the paper the analysis
invokes momentum states and permutation loops
for wave function symmetrization.
These are the basis of
the author's formulation of quantum statistical mechanics
in classical phase space (Attard 2018, 2021).
The reader may be assured that for the ideal gas these give the same results
as obtained originally by London (1938).

Section~\ref{Sec:IntBosons}
generalizes the ideal gas treatment,
including the separation of the ground momentum state
from the excited momentum continuum,
to bosons interacting with arbitrary pair and many-body potentials.
Monte Carlo simulation results show
the dependence upon the interaction potential
of the heat capacity and the $\lambda$-transition temperature.
A mean field theory based on pure permutation loops is developed
and results are given for ground momentum state occupancy
using input from Monte Carlo simulations of Lennard-Jones helium-4.
The resultant heat capacity in the vicinity of the $\lambda$-transition
is shown  in Fig.~\ref{Fig:CvLJ}
and is discussed in section~\ref{Sec:UCv} below.

\begin{figure}[t!]
\centerline{ \resizebox{8cm}{!}{ \includegraphics*{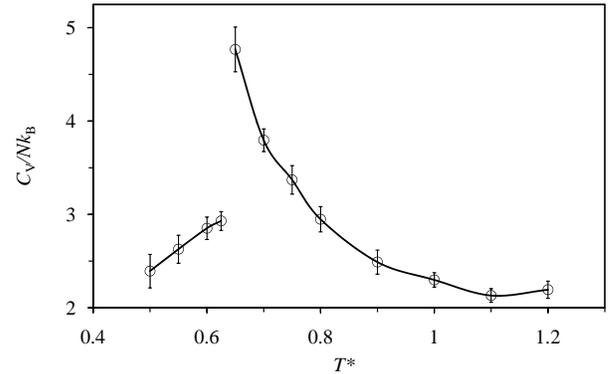} } }
\caption{\label{Fig:CvLJ}
The $\lambda$-transition in a Lennard-Jones fluid.
The specific heat is obtained by Monte Carlo simulation
using pentamer permutation loops.
Below the condensation temperature
the ground momentum state occupancy
is obtained from the pure permutation loop approximation, \S~\ref{Sec:NoMix}.
Note that for  $^4$He, $T[K] = 10.22T^*$.
}
\end{figure}

Section~\ref{Sec:Mixed} analyzes mixed permutation loops.
Monte Carlo dimer results show that the occupancy
of the ground momentum state occurs discontinuously
at the $\lambda$-transition,
which is confirmed by results for higher order mixed loops.
The discontinuity implies a latent heat.
In section~\ref{Sec:Dressed}
dressed ground momentum state bosons are formulated
in a type of ideal solution theory
that locates the condensation transition in Fig.~\ref{Fig:CvLJ}.

The present paper extends the analysis of
the $\lambda$-transition and Bose-Einstein condensation
given in chapter 5 of Attard (2021).
That chapter gives the computational details
and a  molecular-level discussion of the physical nature
of the $\lambda$-transition and superfluidity.

%
\section{Ideal Gas} \label{Sec:Ideal}
\setcounter{equation}{0} \setcounter{subsubsection}{0}
%

In general the grand potential of a quantum system
may be expressed as a series of permutation loop grand potentials
Attard (2018, 2021),
\begin{equation}
\Omega(z,V,T) = \sum_{l=1}^\infty \Omega^{(l)}(z,V,T)  .
\end{equation}
Here $z = e^{\beta \mu}$ is the fugacity,
$V=L^3$ is the volume, and $T = 1/k_\mathrm{B}\beta$ is the temperature.
The monomer $l=1$ contribution to this is the classical contribution.

The ideal gas loop grand potential is exactly
\begin{eqnarray} \label{Eq:ExactIdGas}
- \beta \Omega^{(l)}
& = &
\frac{z^l}{l}
\sum_{n_x,n_y,n_z}
e^{ - l \beta  n^2 \Delta_p^2 /2m} ,
\;\; l \ge 1 .
\end{eqnarray}
Here
the spacing of momentum states is $\Delta_p=2\pi\hbar/L$
(Messiah 1961, Merzbacher 1970).
The boson mass is $m$,
and $\hbar$ is Planck's constant divided by $2\pi$.

Now transform the sum over states to a continuum integral.
In order to analyze the approximation introduced by London (1938),
we need to retain the discrete ground momentum state explicitly.
Therefore to correct for double counting,
we need to subtract the integral over the ground momentum state interval.
This gives
\begin{eqnarray}
- \beta \Omega^{(l)}
& = &
\frac{z^l}{l} \left[
1 +
\Delta_p^{-3} \int
\mathrm{d}{\bf p}  \;
e^{ - l \beta p^2 /2m }
\right. \nonumber \\ && \mbox{ } \left.
-
\Delta_p^{-3} \int_{\Delta_p^3} \mathrm{d}{\bf p} \;
e^{ - l \beta p^2 /2m }
\right]
\nonumber \\ & = &
\frac{z^l}{l}
\left[
1 +
\left(\frac{L^2}{l\Lambda^2} \right)^{3/2}
\right. \nonumber \\ && \mbox{ } \left.
-
\left(\frac{L^2}{l\Lambda^2} \right)^{3/2} \;
\mbox{erf}\!\left( \sqrt{\frac{2\pi l\Lambda^2}{8 L^2}  } \right)^3
\right] .
\end{eqnarray}
The thermal wave length is $\Lambda = \sqrt{2\pi \hbar^2 \beta/m}$.

The asymptotic forms of the error function are
$\mbox{erf}(x) \sim ({2}/{\surd \pi}) x$, $x \rightarrow 0$,
and
$\mbox{erf}(x) \sim 1 - {e^{-x^2}}/{\sqrt{\pi}\,x } $,
$x \rightarrow \infty $
(Abramowitz and Stegun 1970).
It follows that the limiting results for the loop grand potential are
\begin{eqnarray}
- \beta \Omega^{(l)}
& \sim &
\left\{ \begin{array}{ll}
\displaystyle
\frac{ z^l }{l}
\left(\frac{L^2}{l\Lambda^2} \right)^{3/2} , & l \Lambda^2/L^2 \ll 1 ,\\
\displaystyle \rule{0mm}{7mm}
\frac{ z^l }{l} , & l \Lambda^2/L^2 \gg 1.
\end{array} \right.
\end{eqnarray}
In the limit of small argument, $ l \Lambda^2/L^2 \ll 1$
(high temperature, small loops),
the two ground state contributions cancel exactly and
only the integral over excited states survives.
In the limit of large argument, $ l \Lambda^2/L^2 \gg 1$
(low temperature, large loops),
the integral over excited states cancels with
the integral about the ground state,
and only the discrete ground state contribution survives.

Each of these two asymptotic limits
dominates the other in its respective regime.
Hence if we simply add them together
the resultant function is guaranteed correct in both asymptotic limits,
\begin{eqnarray} \label{Eq:LondonIdGas}
- \beta \Omega^{(l)}
& \approx &
\frac{z^l}{l}
+
\frac{z^l}{l}
\left(\frac{L^2}{l\Lambda^2} \right)^{3/2}
\nonumber \\ & \equiv &
- \beta \Omega^{(l)}_0 - \beta \Omega^{(l)}_*  .
\end{eqnarray}
When summed over $l$ this gives
\begin{eqnarray} \label{Eq:OmegaIdGas}
- \beta \Omega
& = &
-\ln [ 1 - z ]  + \Lambda^{-3} V \sum_{l=1}^\infty z^l l^{-3/2}
\nonumber \\ & \equiv &
- \beta \Omega_0 - \beta \Omega_*  .
\end{eqnarray}
This is the expression given by London (1938)
(see also Pathria (1972 chapter~7)).
The first term is the discrete ground state contribution
and the second term is the continuum excited state contribution.
The contribution from the ground state correction
integral about the origin has been neglected.
One sees that the approximate expression
gives  the correct  result in the two respective asymptotic limits.
It might therefore be expected to be a reasonable approximation throughout.

The conventional derivation (c.f.\ Pathria 1972 chapter~7)
asserts that the discrete ground energy state
must be added explicitly to the continuum over the excited energy states
because the volume element for energy,
$\mathrm{d} \varepsilon\, \varepsilon^{1/2}
\propto \mathrm{d} p\, p^2 $,
vanishes at zero and therefore the integral does not include any contribution
from the ground state.
But this is mathematical nonsense,
since the continuum integral includes contributions from the neighborhood
of the origin, $|p_x |\le \Delta_p/2$ etc.,
and it is double counting to include the discrete ground state
in addition to the continuum integral.
The real justification for London's (1938) approximation
is that it is correct in the two asymptotic limits,
and therefore one can hope that it remains accurate
throughout its entire domain.

\begin{figure}[t!]
\centerline{ \resizebox{8cm}{!}{ \includegraphics*{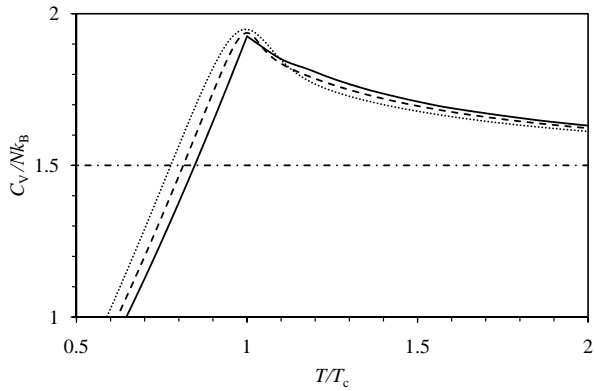} } }
\caption{\label{Fig:CvId}
Specific heat of ideal bosons.
The solid curve uses the London (1938) approximation,
Eq.~(\ref{Eq:LondonIdGas})
(discrete ground state plus continuum excited states),
and the remaining curves use the exact summation
over all momentum states, Eq.~(\ref{Eq:ExactIdGas}).
The density is fixed at $\rho = 0.3$
with $N=500$ (dotted) or $N=5000$ (short dashed).
The condensation temperature $T_\mathrm{c}$
was defined by the location of the maximum for each curve.
The dash-dotted line is the classical result.
Data from Attard (2021).
}
\end{figure}

The London (1938) approximation is tested against exact benchmarks
in Fig.~\ref{Fig:CvId} (Attard 2021 sections~5.1 and 5.2).
One can see that the approximation is very good,
and that it gets better with increasing system size.
One might speculate that it would become exact in the thermodynamic limit.

Unfortunately one cannot so easily check the analogous formulation
for interacting systems.
One should bear this point in mind
in judging the following results for interacting bosons.
These also invoke the London (1938) approximation
of simply adding the discrete ground momentum state
to the continuum integral over the excited momentum states
with no correction for double counting.

%
\section{Interacting Bosons} \label{Sec:IntBosons}
\setcounter{equation}{0} \setcounter{subsubsection}{0}
%

\subsection{Formal Analysis}

Consider a system of $N$  identical bosons
interacting with potential energy $U({\bf q}^N)$,
or more simply $U({\bf q})$,
which does not depend on the momentum state of the bosons.
At any instant there are $N_0$ bosons in the ground momentum state
and $N_*$ in excited momentum states, with $N=N_0+N_*$.
The notation
$j \in N_0$ labels a ground momentum state boson
and
$j \in N_*$ labels an excited momentum state boson.
The momentum eigenvalue of boson $j$ is ${\bf p}_j$.
The classical kinetic energy is
${\cal K}({\bf p}^N) = {\cal K}({\bf p}^{N_*})
= \sum_{j}^{N_*} p_j^2/2m = p^2/2m$.
The normalized  momentum eigenfunctions for the discrete momentum case
are
$|{\bf p}\rangle = V^{-N/2} e^{-{\bf q}\cdot{\bf p}/\mathrm{i}\hbar}$.
The system has volume $V=L^3$
and the spacing between momentum states is
$\Delta_p = 2\pi \hbar/L$ (Messiah 1961, Merzbacher 1970).

The grand partition function for bosons is (Attard 2018, 2021)
\begin{eqnarray}
\Xi
& = &
\mbox{TR}' \; e^{-\beta \hat{\cal H}}
\nonumber \\ & = &
\sum_{N=0}^\infty \frac{z^N}{N!}
\sum_{\hat{\mathrm P}} \sum_{\bf p}
\left\langle \hat{\mathrm P} {\bf p}
\left| e^{-\beta \hat{\cal H}} \right| {\bf p} \right\rangle
\nonumber \\ & = &
\sum_{N=0}^\infty \frac{z^N}{N!}
\sum_{\hat{\mathrm P}} \sum_{\bf p}
\int \mathrm{d}{\bf q}\;
\left\langle \hat{\mathrm P} {\bf p} | {\bf q} \right\rangle \,
\left\langle {\bf q}  \left| e^{-\beta \hat{\cal H}}
\right| {\bf p} \right\rangle
\nonumber \\ & \approx &
\sum_{N=0}^\infty \frac{z^N}{N! V^{N}}
\sum_{\hat{\mathrm P}} \sum_{\bf p}
\int \mathrm{d}{\bf q}\; e^{-\beta {\cal H}({\bf q},{\bf p})}
\frac{
\left\langle \hat{\mathrm P} {\bf p} | {\bf q} \right\rangle
}{ \left\langle {\bf p}  | {\bf q} \right\rangle }
\nonumber \\ & = &
\sum_{N=0}^\infty \frac{z^N}{N! V^{N}}
\sum_{\bf p} \int \mathrm{d}{\bf q}\;
e^{-\beta {\cal H}({\bf q},{\bf p})}
\eta({\bf q},{\bf p}) .
\end{eqnarray}
The permutation operator is $\hat{\mathrm P}$.
In the penultimate equality the commutation function has been neglected;
the error introduced by this approximation is negligible
in systems that are dominated by long range effects (Attard 2018,  2021).
The classical Hamiltonian phase space function is
${\cal H}({\bf q},{\bf p}) = {\cal K}({\bf p}) + U({\bf q})$.
The symmetrization function $\eta({\bf q},{\bf p})$
is the sum of Fourier factors over all boson permutations.

We shall now invoke the London (1938) approximation,
namely transform to the momentum continuum,
include the ground momentum state separately,
and neglect the continuum integral correction
around the momentum ground state.
This gives
\begin{eqnarray}
\Xi
& = &
\sum_{N=0}^\infty \frac{z^N}{N! V^{N}}
\prod_{n=1}^N
\left\{ \delta_{{\bf p}_n,{\bf 0}}
+ \Delta_p^{-3} \int \mathrm{d}{\bf p}_n\; \right\}
\nonumber \\ && \mbox{ } \times
\int \mathrm{d}{\bf q}^N\;
e^{-\beta {\cal H}({\bf q}^N,{\bf p}^N)} \eta({\bf q}^N,{\bf p}^N)
\nonumber \\ & = &
\sum_{N=0}^\infty \frac{z^N}{N! V^{N}}
\sum_{N_0=0}^N \frac{N! \,\Delta_p^{-3N_*}}{N_0!N_*!}
\int \mathrm{d}{\bf p}^{N_*} \;
e^{-\beta {\cal K}({\bf p}^{N_*})}
\nonumber \\ && \mbox{ } \times
\int \mathrm{d}{\bf q}^N\;
e^{-\beta U({\bf q}^N)} \eta({\bf q}^N,{\bf p}^N) .
\end{eqnarray}
In the final equality there are $N_0$ ground momentum state bosons
and $N_* = N-N_0$ excited momentum state bosons;
one could write instead  $\sum_{N,N_0} \Rightarrow \sum_{N_0,N_*} $.
All $N$ bosons contribute to the potential energy
and to the symmetrization function.
Also ${\bf p}_j = {\bf 0}$ if $j \in N_0$.

\subsection{Pure Loops} \label{Sec:NoMix}

The symmetrization function is the sum of all permutations
of the bosons in the Fourier factors.
Each permutation can be expressed as a product of disjoint permutation loops
(Attard 2018, 2021).
In the present context
permutation loops can be classified as pure or mixed.
A pure permutation loop consists only of ground momentum state bosons,
or only of excited  momentum state bosons.
A mixed permutation loop contains both.

In this section we shall pursue the leading order approximation
in which all mixed permutation loops are neglected,
\begin{equation}
\eta({\bf q}^N,{\bf p}^N)
\approx
\eta_0({\bf q}^{N_0},{\bf p}^{N_0})
\eta_*({\bf q}^{N_*},{\bf p}^{N_*}) .
\end{equation}
Here $\eta_0$ is the sum total of weighted permutations
of ground momentum state bosons.
Since their momentum is zero, ${\bf p}_j = {\bf 0}$, $j \in N_0$,
one has that
\begin{equation}
\frac{
\left\langle \hat{\mathrm P} {\bf p}^{N_0} | {\bf q}^{N_0} \right\rangle
}{ \left\langle {\bf p}^{N_0}  | {\bf q}^{N_0} \right\rangle }
= 1 ,
\mbox{ all } \hat{\mathrm P}.
\end{equation}
Hence
\begin{equation} \label{Eq:N0!}
\eta_0({\bf q}^{N_0},{\bf p}^{N_0}) = N_0!.
\end{equation}
Of relevance to the later discussion of superfluidity,
this result is a manifestation of quantum non-locality
for ground momentum state bosons.
All the ground momentum state bosons in the system contribute equally
regardless of spatial location.

The remaining factor
$\eta_*$ is the sum total of weighted permutations
of excited momentum state bosons.
In previous work (Attard 2021)
the number of discrete ground momentum state bosons
was taken to be negligible, $N_0=0$
and the continuum integral over momentum space
was applied to all $N$ bosons.
Hence all of those previous formal results carry over directly
to the present case of pure excited momentum state permutation loops
with the replacement $N \Rightarrow N_*$.
(An adjustment has to be made to the statistical averages
when the ground momentum state bosons are not negligible,
as is discussed below.)
In particular, $\eta_*$, is a sum of products of loops.
One can define $\stackrel{\circ}{\eta}\!_*
=\sum_{l\ge2} \eta^{(l)}_* $ as the sum of single loops,
and one can write either
\begin{equation}
\eta_*({\bf q}^{N_*},{\bf p}^{N_*})
=
\exp \stackrel{\circ}{\eta}\!_*({\bf q}^{N_*},{\bf p}^{N_*})  ,
\end{equation}
or else
\begin{equation}
\langle \eta_* \rangle_\mathrm{cl}
= e^{ \langle \stackrel{\circ}{\eta}\!_* \rangle_\mathrm{cl} } .
\end{equation}
One can drop the subscript $*$ on $\eta$
when the bosons that contribute are obvious from its arguments.

It may be helpful to illustrate these two result by writing out
the first few terms of the permutation series explicitly,
\begin{eqnarray}
\lefteqn{
\eta({\bf q}^{N_*},{\bf p}^{N_*})
} \nonumber \\
& = &
1
+ \sum_{j,k}^{N_*} \!''\;
e^{-{\bf p}_{j} \cdot {\bf q}_{jk} /\mathrm{i}\hbar}
e^{-{\bf p}_{k} \cdot {\bf q}_{kj} /\mathrm{i}\hbar}
\nonumber \\ && \mbox{ }
+ \sum_{j,k,n}^{N_*} \!''\;
e^{-{\bf p}_{j} \cdot {\bf q}_{jk} /\mathrm{i}\hbar}
e^{-{\bf p}_{k} \cdot {\bf q}_{kn} /\mathrm{i}\hbar}
e^{-{\bf p}_{n} \cdot {\bf q}_{nj} /\mathrm{i}\hbar}
\nonumber \\ && \mbox{ }
+ \sum_{j,k,n,m}^{N_*} \hspace{-2mm}''\hspace{2mm}
e^{-{\bf p}_{j} \cdot {\bf q}_{jk} /\mathrm{i}\hbar}
e^{-{\bf p}_{k} \cdot {\bf q}_{kj} /\mathrm{i}\hbar}
\nonumber \\ && \hspace{1.8cm} \times
e^{-{\bf p}_{n} \cdot {\bf q}_{nm} /\mathrm{i}\hbar}
e^{-{\bf p}_{m} \cdot {\bf q}_{mn} /\mathrm{i}\hbar}
\nonumber \\ && \mbox{ }
+ \ldots
\nonumber \\ & = &
1
+ \eta^{(2)}({\bf q}^{N_*},{\bf p}^{N_*})
+ \eta^{(3)}({\bf q}^{N_*},{\bf p}^{N_*})
\nonumber \\ && \mbox{ }
+ \frac{1}{2}
\eta^{(2)}({\bf q}^{N_*},{\bf p}^{N_*})^2
+ \ldots
\end{eqnarray}
The double prime indicates that the sums are over unique permutations,
and also that in any product term
no boson may belong to more than one permutation loop.
The first term is the monomer or unpermuted one,
and it gives rise to classical statistics.
The second term is the dimer, the third is the trimer,
and the fourth is the double dimer.
The second equality no longer forbids permutation loop intersections;
the error from this approximation ought to be negligible
in the thermodynamic limit.

In view of these results,
in particular the one that writes the classical average
of the full symmetrization function
as the exponential of the classical average
of the single loop symmetrization function,
it is straightforward to write the grand potential
as the sum of loop grand potentials,
\begin{eqnarray}
\Omega(z,V,T)
& = &
-k_\mathrm{B}T \ln \Xi(z,V,T)
\nonumber \\ & = &
\sum_{l=1}^\infty \Omega^{(l)}_*(z,V,T) .
\end{eqnarray}

The monomer grand potential is just the classical grand potential,
with a trivial adjustment for $N_0$ and $N_*$ as independent.
It is given by
\begin{eqnarray} \label{Eq:OmegaCl}
e^{ -\beta \Omega^{(1)}(z,V,T) }
& = &
\Xi_\mathrm{cl}(z,V,T)
\nonumber \\ & = &
\sum_{N_0,N_*}
\frac{ z^N\Delta_p^{-3N_*}}{N_*!V^{N}}
\int \mathrm{d}{\bf p}^{N_*} \;
e^{-\beta {\cal K}({\bf p}^{N_*})}
\nonumber \\ && \mbox{ } \times
\int \mathrm{d}{\bf q}^N\;
e^{-\beta U({\bf q}^N)}
\nonumber \\ & = &
\sum_{N_0,N_*}
\frac{ z^N\Lambda^{-3N_*}}{N_*!V^{N_0}}
\int \mathrm{d}{\bf q}^N\;
e^{-\beta U({\bf q}^N)}
\nonumber \\ & = &
\sum_{N_0,N_*}
\frac{ z^N\Lambda^{-3N_*}}{N_*!V^{N_0}}
Q(N,V,T).
\end{eqnarray}
Again $N=N_0+N_*$.
The $N_0!$ in the denominator has canceled with $\eta_0=N_0!$
in the integrand,
since this multiplies each of the terms in $\eta_*$.
The classical configurational integral, $Q(N,V,T)$,
does not distinguish
between ground and excited momentum state bosons.
The thermal wave length is $\Lambda = \sqrt{2\pi\beta\hbar^2/m}$.

The loop grand potentials $l \ge 2$ are classical averages
(Attard 2021 section~5.3),
which can be taken in a canonical system
\begin{eqnarray} \label{Eq:OmegaPure}
-\beta \Omega^{(l)}_*
& = &
\left\langle \eta^{(l)}
({\bf p}^{N_*},{\bf q}^{N_*})
\right\rangle_{N_0,N_*,\mathrm{cl}}
\nonumber \\ & = &
\left\langle G^{(l)}({\bf q}^{N_*})
\right\rangle_{N_0,N_*,\mathrm{cl}}
\nonumber \\ & = &
\left(\frac{N_*}{N}\right)^{l}
\left\langle G^{(l)}({\bf q}^{N}) \right\rangle_{N,\mathrm{cl}}
\nonumber \\ & \equiv &
N_* \left(\frac{N_*}{N}\right)^{l-1} g^{(l)}  .
\end{eqnarray}
We have transformed the average from the mixed $\{N_0,N_*\}$ system
to the classical configurational system of $N$ bosons
that does not distinguish their state.
This transformation invokes a factor of  $\left({N_*}/{N}\right)^{l}$,
which is the uncorrelated probability that $l$ bosons
chosen at random in the original mixed system are all excited.
The Gaussian position loop function is
\begin{equation}
G^{(l)}({\bf q}^{N})
=
\sum_{j_1,\ldots,j_l}^N\hspace{-.2cm}'\hspace{.1cm}
e^{-\pi q_{j_l,j_1}^2 /\Lambda^2 }
\prod_{k=1}^{l-1}
e^{-\pi q_{j_k,j_{k+1}}^2 /\Lambda^2 }  .
\end{equation}
The prime indicates that no two indeces may be equal
and that distinct loops must be counted once only.
There are $N!/(N-l)!l$ distinct $l$-loops here,
the overwhelming number of which are negligible upon averaging.
Since the pure excited momentum state permutation loops
are compact in configuration space,
one can define an intensive form
of the average loop Gaussian,
$g^{(l)} \equiv
\left\langle G^{(l)}({\bf q}^{N}) \right\rangle_{N,\mathrm{cl}} /N$.
This is convenient because it does not depend upon $N_*$.

The mix of ground and excited momentum state bosons
can be determined by minimizing the grand potential
with respect to $N_*$ at constant $N$.
This is equivalent to maximizing the total entropy.
The derivative is
\begin{equation} \label{Eq:dW/dN*}
\left( \frac{\partial (-\beta \Omega) }{\partial N_*} \right)_N
=
-\ln \frac{N_*  \Lambda^{3 }}{ V }
+
\sum_{l=2}^\infty  l\left(\frac{N_*}{N}\right)^{l-1} g^{(l)} .
\end{equation}
If this is positive, then $N_*$ should be increased,
and \emph{vice versa}.
In terms of the number density $\rho_* = N_*/V$ and $\rho = N/V$,
if $\rho_* \Lambda^3 \le \rho \Lambda^3 <  1$,
then both terms are positive
(since $g^{(l)}\ge 0$).
In this case the only stable solution is $\overline \rho_* = \rho$.
This is the known result on the high temperature side
of the $\lambda$-transition.

\subsection{Pure Loop Numerical Results}

\begin{figure}[t!]
\centerline{ \resizebox{8cm}{!}{ \includegraphics*{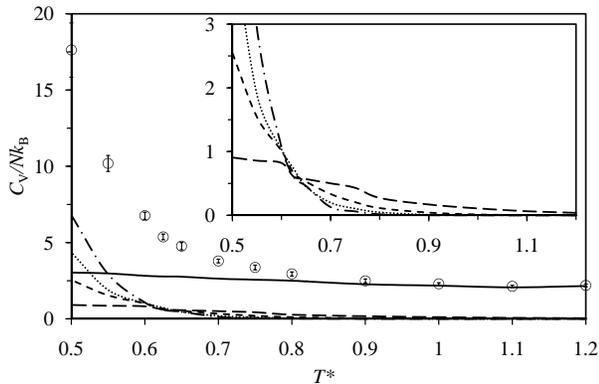} } }
\caption{\label{Fig:Cv}
Canonical Monte Carlo (homogeneous, $N_*=N=5,000$)
results  for the
specific heat capacity for Lennard-Jones $^4$He
along the saturation curve.
The symbols are the total of the monomer and loop contributions.
The full curve is the monomer or classical contribution.
The loop contribution curves are
dimer (long dashed),
trimer (short dashed),
tetramer (dotted),
and pentamer (dash-dotted).
Each arm of an error bar is twice the standard error.
\textbf{Inset.} Magnification of the loop contributions.
}
\end{figure}

Figure~\ref{Fig:Cv} shows the simulated heat capacity
for interacting bosons,
namely helium-4 with Lennard-Jones pair interactions.
Details of the Monte Carlo algorithm are given by Attard (2021 section~5.3.2).
The difference between the present results
and those reported by Attard (2021 section~5.3.3)
is that the earlier results were for an inhomogeneous canonical system
consisting of a liquid drop in equilibrium with its own vapor.
The present results are for a homogenous canonical system
at the saturated liquid density
previously established at each temperature by the inhomogeneous  simulation.
Artifacts due to the fluctuations in the liquid drop
that contribute significantly to the heat capacity for the total system
reported by  Attard (2021)
are absent in Fig.~\ref{Fig:Cv}.
It can be seen that the magnitude of the heat capacity
for the Lennard-Jones liquid
is much larger than that of the ideal gas, Fig.~\ref{Fig:CvId},
and comparable to that measured experimentally for $^4$He
in the vicinity of the $\lambda$-transition
(Donnelly  and  Barenghi  1998).

The individual loop contributions come from expressing
the second temperature derivative of the loop grand potentials
as classical averages (Attard 2021 section~5.3.1).
It can be seen that for  $T* \agt 0.9$
the quantum contributions are negligible.
For $0.6 \alt T^* \alt 0.8 $
the loop contributions increase with decreasing temperature,
with the lower order loops having largest effect.
At about $T^*=0.6$ the order of the loops flips,
and the loop series appears divergent.
This suggests that for the Lennard-Jones fluid
the $\lambda$-transition occurs around  about $T^*_\mathrm{c} \approx 0.6$,
which for the usual Lennard-Jones parameters of helium-4
(van Sciver 2012) corresponds to $T_\mathrm{c} \approx 6$\,K.
(The temperature $T^* \approx 0.6$
also signals the onset of ground momentum state condensation,
as shown by the results in sections~\ref{Sec:SingExLoop}
and \ref{Sec:Dressed} below.)

One should not be unduly concerned that this predicted
$\lambda$-transition temperature differs from the measured one,
$T_\mathrm{c} =$2.17\,K (Donnelly  and  Barenghi  1998).
The present Lennard-Jones parameters,
$ \varepsilon_\mathrm{He} = 10.22 k_\mathrm{B}$\,J
and $\sigma_\mathrm{He} = 0.2556$\,nm
(van Sciver 2012)
could be simply rescaled,
$ \varepsilon = 0.43 \varepsilon_\mathrm{He}$
and $\sigma = 1.4 \sigma_\mathrm{He}$,
to bring the transition temperature
and the saturation density into consonance with the measured values.
The functional form of the Lennard-Jones pair potential,
and the absence of many-body potentials,
appear inadequate for a quantitatively accurate description
of liquid helium at these low temperatures.
The real message from Fig.~\ref{Fig:Cv}
is that particle interactions play an essential role
in the $\lambda$-transition
and that this can be explored with computer simulations.

A second limitation of the simulation results in
Fig.~\ref{Fig:Cv} is that they used classical canonical  averages
with $N$ bosons and the momentum continuum,
$\langle \ldots \rangle_{N,\mathrm{cl}}$.
This is valid if the number of bosons in the ground momentum state
is negligible compared to the number of excited momentum state bosons,
$N_0 \ll N_*$.
Such an assumption obviously breaks down
after the $\lambda$-transition.
In the unmixed case the primary correction for this effect
is to multiply the energy due to an $l$-loop,
$l \ge 2$,  by $(\overline N_*/N)^l$,
using an estimate of the optimum fraction of excited momentum state bosons
at each temperature.
This is done below the $\lambda$-transition in Fig.~\ref{Fig:CvLJ}
and is discussed in \S\ref{Sec:UCv}.
If condensation increases continuously with decreasing temperature,
as will now be shown in the unmixed case,
such a correction reduces the heat capacity in Fig.~\ref{Fig:Cv}
following the $\lambda$-transition.

\begin{table}[tb]
\caption{ \label{Tab:gloop}
Intensive Gaussian loop
$ g^{(l)} = \langle G^{(l)} \rangle_{N,\mathrm{cl}}/N$
from liquid drop Monte Carlo simulations
($N=500$, $\rho \sigma^3=0.3$)
for saturated Lennard-Jones He$^4$ at various temperatures,
$T^* \equiv k_\mathrm{B} T /\varepsilon$.
}
\begin{center}
\begin{tabular}{c c c c c }
\hline\noalign{\smallskip}
$l$ & $ T^*= 1 $ & 0.9 & 0.8 & 0.7 \\
\hline \rule{0cm}{0.4cm}%
2 & 2.44E-03 & 5.59E-03 & 1.24E-02 & 2.70E-02 \\
3 & 6.77E-05 & 2.81E-04 & 1.11E-03 & 4.25E-03 \\
4 & 3.84E-06 & 2.60E-05 & 1.67E-04 & 1.06E-03 \\
5 & 3.20E-07 & 3.28E-06 & 3.38E-05 & 3.57E-04 \\
$l$  & $ T^*= 0.6 $ & 0.55 & 0.5 & 0.45 \rule{0cm}{0.4cm}\\
\hline \rule{0cm}{0.4cm}%
2 & 5.90E-02 & 8.72E-02 & 1.29E-01 & 1.96E-01 \\
3 & 1.63E-02 & 3.20E-02 & 6.32E-02 & 1.33E-01 \\
4 & 6.62E-03 & 1.68E-02 & 4.28E-02 & 1.30E-01 \\
5 & 3.62E-03 & 1.19E-02 & 3.94E-02 & 1.56E-01 \\
6 & 2.08E-03 & 8.76E-03 & 3.72E-02 & 2.24E-01 \\
\hline
\end{tabular} \\
\end{center}
\end{table}

Using the mean field results of the previous section
the number of ground momentum state bosons can be estimated.
Table~\ref{Tab:gloop} gives the intensive Gaussian loop weights,
$ g^{(l)} \equiv
\langle G^{(l)}({\bf q}^{N}) \rangle_{N,\mathrm{cl}} /N$.
These were obtained from the Monte Carlo simulations of
Lennard-Jones $^4$He (Attard 2021).
The statistical error was less than the displayed digits.
It can be seen that the terms increase with decreasing temperature.
At each temperature the terms decrease
with increasing loop size, except possibly at the lowest temperature shown.

\begin{table}[tb]
\caption{ \label{Tab:dens}
Saturation density and
fraction of ground momentum state bosons
in Lennard-Jones He$^4$.
From liquid drop Monte Carlo simulations with $N_*=N=500$
at an overall density of $\rho \sigma^3=0.3$.
}
\begin{center}
\begin{tabular}{c c c c c c c}
\hline\noalign{\smallskip}
$k_\mathrm{B} T /\varepsilon $ & $\rho \sigma^3$  &  $\rho \Lambda^3$  &
$T$ (K)$^\dag$  & $\rho_\mathrm{m}$ (kg\,m$\!^{-3}$)$^\dag$  &
$\overline N_0/N$   \\
\hline 
1    & 0.62 & 0.76 & 10.220 & 246.76 & 0 \\ 
0.9  & 0.73 & 1.04 & 9.198  & 290.54 & 0.027 \\ 
0.8  & 0.77 & 1.31 & 8.176   & 306.46 & 0.220 \\ 
0.7  & 0.81 & 1.68 & 7.154  & 322.38 & 0.381 \\ 
0.6  & 0.88 & 2.31 &6.132   & 350.24 & 0.536 \\ 
0.55 &  0.87 & 2.60 & 5.621 & 346.26 & 0.583 \\ 
0.5  & 0.91 & 3.13 & 5.110  & 366.16 & 0.635 \\ 
0.45 & 0.95 & 3.82 & 4.599  & 376.91 & 0.682 \\
0.4  & 0.98 & 4.72 & 4.088  & 390.04 & 0.709 \\ 
\hline
\end{tabular} \\
$^\dag$$ \varepsilon_\mathrm{He} = 10.22 k_\mathrm{B}$\,J
and $\sigma_\mathrm{He} = 0.2556$\,nm.
\end{center}
\end{table}

Table~\ref{Tab:dens} shows the optimal fraction of bosons
in the system that are in the ground momentum state
as a function of temperature.
The mean field theory, Eq.~(\ref{Eq:dW/dN*}), was used
together with the results in Table~\ref{Tab:gloop}.
The results represent the  stable solutions.
At $T^* = 0.45$, changing $l_\mathrm{max}$ from 6 to 5
changes $\overline N_0/N$ from 0.682 to 0.684.

One can compare these results
to the exact calculations for the ideal gas,
Eq.~(\ref{Eq:ExactIdGas}).
As reported in Attard (2021 section~5.2)
for $N=500$ about 20\% of the bosons are in the ground momentum state
at the $\lambda$-transition.
For $N=1000$ it is about 10\%,
and for $N=5000$ it is about 5\%.
In the thermodynamic limit
the ideal gas apparently shows a continuous condensation transition,
beginning with zero ground momentum state occupation
at the transition itself,
followed by a continuous increase in occupancy below the transition.
In contrast, the present results
for the Lennard-Jones liquid, Table~\ref{Tab:dens},
have a substantial fraction of bosons
occupying the ground momentum state at the transition.
The present mean field results indicate
that the ground momentum state occupancy
increases continuously through the transition.
In the light of the following results
for mixed permutation loops,
this continuity appears to be an artefact of the no mixing approximation,
whereas the non-zero value for ground momentum state occupancy
\emph{at} the condensation  transition appears to be reliable.

\comment{ 
One might speculate
that excluding mixed loops
precludes any coupling of ground and excited momentum state bosons.
Since the pair correlation function in a mixed system
would have a larger peak
for excited-excited bosons than for ground-excited bosons,
(i.e.\ the density in the vicinity of an excited momentum state boson
is greater than that in the vicinity of a ground momentum state boson),
the uncorrelated probability $(N_*/N)^{l-1}$
should be replaced by something larger.
In this case one would require $\rho^* \Lambda^3$
to be significantly greater than unity
in Eq.~(\ref{Eq:dW/dN*}) before condensation into
the momentum ground state would occur.
Such speculation  remains to be confirmed
by actual calculations or simulations.
} 

%
\section{Mixed Loops} \label{Sec:Mixed}
\setcounter{equation}{0} \setcounter{subsubsection}{0}
%

In order to go beyond the mean field (i.e.\  no mixing) approach
one must include mixed permutation loops
that contain both ground and excited momentum state bosons.
Such mixed loops are essential because they provide
the nucleating mechanism for ground momentum state condensation.
Despite the strong inference
based on the experimental and computational evidence
that mixed loops nucleate condensation,
their very existence poses
a surprising conceptual and mathematical challenge.

For the ideal gas, only pure permutation loops are allowed:
momentum states cannot be mixed in any one loop.
The prohibition on mixing in the ideal gas
is due to the orthogonality of the momentum eigenfunctions;
the ideal gas configuration integral is the same
as the expectation value integral.
For interacting bosons this restriction does not  apply directly
because the interaction potential in the Maxwell-Boltzmann factor
means that the configuration integral differs
from the expectation value integral in which orthogonality
would otherwise hold.

This observation might lead one to conclude
that mixed ground and excited momentum permutation loops
for interacting bosons are fully permitted.
But closer inspection reveals a complication,
namely that at large separations the interaction potential goes to zero,
which means that the bosons behave ideally with respect to each other.
From this point of view mixed ground and excited momentum state loops
would be forbidden also for interacting bosons
because the large separation regime
dominates the configuration integral.
There is an issue to be resolved here
that turns on the form of the London (1938) approximation,
the nature of generalized functions, and the order of integration.

In the discrete momentum picture,
\begin{equation}
\int_V \mathrm{d} {\bf q}_{12} \;
e^{-{\bf p}_{12} \cdot {\bf q}_{12} /\mathrm{i}\hbar}
=
V \delta_{{\bf p}_{1},{\bf p}_{2} },
\end{equation}
whereas in the continuum momentum picture
\begin{equation}
\int \mathrm{d} {\bf q}_{12} \;
e^{-{\bf p}_{12} \cdot {\bf q}_{12} /\mathrm{i}\hbar}
=
(2\pi\hbar)^3 \delta({\bf p}_{1}-{\bf p}_{2} ).
\end{equation}
In these Kronecker and Dirac $\delta$-functions respectively appear.
Because of this difference,
one can only apply the London approximation \emph{after}
performing the position integral
for the particle density asymptote, as will be demonstrated.

\subsection{Mixed Dimer} \label{Sec:MixDim}

In order to account for mixed permutation loops
write the symmetrization function as
\begin{eqnarray}
\eta({\bf q},{\bf p})
& = &
\eta_0({\bf q}^{N_0},{\bf p}^{N_0}) \eta_*({\bf q}^{N_*},{\bf p}^{N_*})
 \nonumber \\ && \mbox{ } \times
\big[ 1 + \eta_\mathrm{mix}({\bf q}, {\bf p}) \big] .
\end{eqnarray}
Here $\eta_0=N_0!$
is the pure ground momentum state symmetrization function,
$ \eta_*$
is the pure  excited momentum state  symmetrization function,
and $\eta_\mathrm{mix}$ is the sum of mixed loop weights.

The ground momentum state bosons in any one mixed loop
are inaccessible to $\eta_0$ and their number have to be removed from it.
This is most easily done by fixing $\eta_0=N_0!$
and multiplying each mixed symmetrization loop with $n_0$
ground momentum state bosons by $(N_0-n_0)!/N_0!$.
The excited loops are compact,
and so we can neglect any similar dependence on $n_*$
in the thermodynamic limit.

Here we focus on the leading mixed contribution,
which is the mixed dimer,
\begin{equation} \label{Eq:eta2MixPhase}
\eta^{(1;1)}_{0*}({\bf q},{\bf p})
=
\sum_{j}^{N_0}  \sum_{k}^{N_*}
 e^{-{\bf p}_{k} \cdot {\bf q}_{kj}/\mathrm{i}\hbar} .
\end{equation}
This  is also the dimer chain, $\tilde \eta^{(2)}$,
and it belongs to the class of singly excited mixed loops, $\eta^{(n;1)}$,
both of which are treated below.
As mentioned, we can account  for the fact that there are now $N_0-1$
ground momentum state bosons available for pure loops
by dividing this by $N_0$,
which will then  leave  $\eta_0=N_0!$ unchanged.

We suppose that we can factorize the classical average
of the symmetrization function
\begin{equation}
\left\langle
\eta_0 \eta_* \eta^{(1;1)}_{0*} /N_0
 \right\rangle_{z,\mathrm{cl}}
 =
\left\langle \eta_0 \eta_*  \right\rangle_{z,\mathrm{cl}} \;
\left\langle \eta^{(1;1)}_{0*} /N_0  \right\rangle_{z,\mathrm{cl}} .
\end{equation}
The unmixed factor was dealt with above.

Using for convenience a canonical rather than a grand canonical average,
in the discrete momentum case one can write for this
\begin{eqnarray}
\lefteqn{
\left\langle \eta^{(1;1)}_{0*} /N_0 \right\rangle_{N_0,N_*,\mathrm{cl}}
} \nonumber \\
& = &
\frac{\Lambda^{3N_*}}{V^{N_*}}
\prod_j^{N_0} \delta_{{\bf p}_j,{\bf 0}}
\prod_k^{N_*} \sum_{{\bf p}_k}\!^{(p_k>0)} e^{-\beta p_k^2/2m }
 \nonumber \\ & & \mbox{ } \times
\left\langle
\frac{1}{N_0}
\sum_j^{N_0} \sum_k^{N_*}
 e^{-{\bf p}_{k} \cdot {\bf q}_{kj}/\mathrm{i}\hbar}
 \right\rangle_{N,\mathrm{cl}}
 \nonumber \\ & = &
\frac{\Lambda^{3N_*}}{V^{N_*}}
\frac{N_*}{N(N-1)}
\prod_j^{N_0} \delta_{{\bf p}_j,{\bf 0}}
\prod_k^{N_*} \sum_{{\bf p}_k}\!^{(p_k>0)} e^{-\beta p_k^2/2m }
 \nonumber \\ & & \mbox{ } \times
\int_V \mathrm{d}{\bf q}_1\,\mathrm{d}{\bf q}_2\,
\rho^{(2)}_N({\bf q}_1,{\bf q}_2)
 e^{-{\bf p}_{1} \cdot {\bf q}_{12}/\mathrm{i}\hbar} .
\end{eqnarray}
The angular brackets represent the configurational average,
the momentum average being explicit:
the prefactor for the first equality
is the normalization for the momentum states,
which invokes the continuum integral
for the excited states because in this case there is no issue
with generalized functions.
The second equality converts the configuration average to an integral
over the canonical pair density,
which is normalized as $\int_V \mathrm{d}{\bf q}_1\,\mathrm{d}{\bf q}_2\,
\rho^{(2)}_N({\bf q}_1,{\bf q}_2)
= N(N-1)$.
In this case particle 1 is in an excited momentum state
and particle 2 is in the momentum ground state,
although this makes no difference to the configuration integral.

The asymptotic contribution to the configurational integral is
\begin{equation}
\rho^2
\int_V \mathrm{d}{\bf q}_1\,\mathrm{d}{\bf q}_2\,
e^{-{\bf p}_{1} \cdot {\bf q}_{12}/\mathrm{i}\hbar}
=
\rho^2 V \delta_{{\bf p}_1,{\bf 0}}
= 0  .
\end{equation}
Since $p_1 > 0$ by design, this vanishes.
Therefore
in order to
transform the sum over excited states to
the integral over the momentum continuum
and interchange the order of integration
we must first replace the pair density by its connected part,
\begin{equation}
\rho^{2} h^{(2)}_N({\bf q}_1,{\bf q}_2)
=
\rho^{(2)}_N({\bf q}_1,{\bf q}_2) - \rho^2 .
\end{equation}
Here we assume a homogeneous system,
$\rho^{(1)}_N({\bf q}_1) = \rho = N/V$,
in which case
the total correlation function
depends upon the particle separation,
$h^{(2)}_N({\bf q}_1,{\bf q}_2) = h^{(2)}_N(q_{12})$.

Subtracting the asymptote
we are left with a short-ranged integrand
and therefore no generalized function.
Hence we may transform to the momentum continuum,
interchange the order of integration,
and perform the integral over the excited states.
These give
\begin{eqnarray}
\lefteqn{
\left\langle \eta^{(1;1)}_{0*} /N_0 \right\rangle_{N_0,N_*,\mathrm{cl}}
} \nonumber \\
& = &
\frac{\Lambda^{3N_*}}{V^{N_*}}
\frac{N_*}{N(N-1)}
\prod_j^{N_0} \delta_{{\bf p}_j,{\bf 0}}
\prod_k^{N_*} \sum_{{\bf p}_k}\!^{(p_k>0)} e^{-\beta p_k^2/2m }
\nonumber \\ && \mbox{ } \times
\int \mathrm{d}{\bf q}_1\,\mathrm{d}{\bf q}_2\,
\rho^2 h^{(2)}_N({\bf q}_1,{\bf q}_2)
 e^{-{\bf p}_{1} \cdot {\bf q}_{12}/\mathrm{i}\hbar}
 \nonumber \\ & = &
\frac{\Lambda^{3N_*}}{V^{N_*}}
\frac{N_*}{N(N-1)} \Delta_p^{-3N_*}
\int \mathrm{d}{\bf p}^{N_*}\; e^{-\beta {\cal K}({\bf p}^{N_*}) }
\nonumber \\ && \mbox{ } \times
\int \mathrm{d}{\bf q}_1\,\mathrm{d}{\bf q}_2\;
\rho^2 h^{(2)}_N({\bf q}_1,{\bf q}_2)
 e^{-{\bf p}_{1} \cdot {\bf q}_{12}/\mathrm{i}\hbar}
 \nonumber \\ & = &
\frac{\rho^2 N_*}{N(N-1)}
\int \mathrm{d}{\bf q}_1\,\mathrm{d}{\bf q}_2\;
h^{(2)}_N({\bf q}_1,{\bf q}_2)
 e^{-\pi q^2_{12}/\Lambda^2 }
 \nonumber \\ & = &
\frac{N_* }{V}
\int \mathrm{d}{\bf q}_{12}\,
h^{(2)}_N(q_{12})  e^{-\pi q_{12}^2 /\Lambda^2} .
\end{eqnarray}
This can also be written
\begin{eqnarray}  \label{Eq:MixDimer}
\lefteqn{
\left\langle \eta^{(1;1)}_{0*}/N_0 \right\rangle_{N_0,N_*,\mathrm{cl}}
} \nonumber \\
& = &
\frac{N_* V}{N^2}
\int \mathrm{d}{\bf q}_{12}\;
\rho^{(2)}_N(q_{12})  e^{-\pi q_{12}^2 /\Lambda^2}
-
\frac{N_* V}{N^2} \rho^2 \Lambda^{3}
\nonumber \\ & = &
\frac{N_*}{N}
\left\langle
\frac{2}{N}
\sum_{j<k}^N
 e^{-\pi q_{jk}^2 /\Lambda^2}
\right\rangle_{N,\mathrm{cl}}
-
\frac{N_* V}{N^2} \rho^2 \Lambda^{3}
\nonumber \\ & = &
\frac{N_*}{N}
\left\langle
\frac{1}{N}
\sum_{j,k}^N\!'
\left[  e^{-\pi q_{jk}^2 /\Lambda^2}
- \frac{\rho \Lambda^3}{N}  \right]
\right\rangle_{N,\mathrm{cl}}
\nonumber \\ & \equiv &
\frac{N_*}{N} \left\langle \tilde \eta^{(2)} /N
\right\rangle_{N,\mathrm{cl}}^\mathrm{corr} .
\end{eqnarray}
The final average, which is in a canonical classical system of $N$ particles,
has the asymptotic correction as defined by the preceding equality.
The result is intensive.

For high temperatures
$\Lambda \alt \sigma$,
which means that
the integral vanishes in the core, $\rho^{(2)}_N(q_{12}) = 0$,
$ q_{12} \alt \sigma$.
Hence at high temperatures
$  \langle  \eta^{(1;1)}_{0*}/N_0 \rangle_{N_0,N_*,\mathrm{cl}}
\sim -\rho \Lambda^3$.
That this is negative means that such mixed dimers are
entropically unfavorable in this regime,
which suppresses ground momentum state occupation,
which will turn out to be significant.

\comment{ 
In terms of the quantity
$\tilde g^{(2)} = N^{-1} \langle
\sum_{j<k}^N e^{-\pi q_{jk}^2/\Lambda^2} \rangle_{N,\mathrm{cl}}$,
one has that
$ \langle \stackrel{\circ}{\eta}\!^{(2)}_{0*} \rangle_{N_0,N_*,\mathrm{cl}}
\approx 2 \tilde g^{(2)} - \rho \Lambda^3$.

The pure ground momentum state dimer is
\begin{eqnarray}
\lefteqn{
\left\langle
\stackrel{\circ}{\eta}\!^{(2)}_{00}
\right\rangle_{N_0,N_*,\mathrm{cl}}
} \nonumber \\
& = &
\frac{\Lambda^{3N_*}}{V^{N_*}}
\prod_j^{N_0} \delta_{{\bf p}_j,{\bf 0}}
\prod_k^{N_*} \sum_{{\bf p}_k}\!^{(p_k>0)} e^{-\beta p_k^2/2m }
 \nonumber \\ & & \mbox{ } \times
\left\langle
\frac{1}{N_0(N_0-1)}
\sum_{j<k}^{N_0}
 e^{-{\bf p}_{jk} \cdot {\bf q}_{jk}/\mathrm{i}\hbar}
 \right\rangle_{N,\mathrm{cl}}
 \nonumber \\ & = &
\frac{1}{2} .
\end{eqnarray}
} 

\subsection{Singularly Exciting Mixed Loops} \label{Sec:SingExLoop}

\subsubsection{Analysis}

Consider mixed $l$-loops with one excited momentum state boson,
labeled 1 or $j_1$, and $l-1$ ground momentum state bosons,
labeled $2,\ldots,l$ or $k_2,\ldots,k_l$,
whose weight we shall denote as $ \eta^{(l-1;1)}_{0*}$.
A particular such loop
with particle 1 excited has symmetrization factor
$e^{-{\bf p}_1 \cdot {\bf q}_{12} /\mathrm{i}\hbar}$,
which is independent of the positions of all the ground momentum state bosons
in the loop,
${\bf q}_3,{\bf q}_4,\ldots,{\bf q}_{l-1}$,
except the adjacent one labeled 2.
Therefore we can arrange them in $(l-2)!$ ways without changing
the value of the symmetrization factor.
The total weight involving all such singly excited mixed loops is
\begin{eqnarray}
\lefteqn{
\eta_{0}\, \eta_* \eta^{(l-1;1)}_{0*}
} \nonumber \\
& = &
(N_0-l+1)! \, \eta_*(N_*)
\sum_{k_2,\ldots,k_l}^{N_0}\!\!\!'\; \sum_{j_1}^{N_*}
e^{-{\bf p}_{j_1} \cdot {\bf q}_{j_1,k_2} /\mathrm{i}\hbar}
\nonumber \\ & = &
(N_0-l+1)! \, \eta_*(N_*)
\frac{(N_0-1)!}{(N_0-l+1)!}
\nonumber \\ && \mbox{ } \times
\sum_{k_2}^{N_0} \sum_{j_1}^{N_*}
e^{-{\bf p}_{j_1} \cdot {\bf q}_{j_1,k_2} /\mathrm{i}\hbar}
\nonumber \\ & = &
\eta_0(N_0)\, \eta_*(N_*)
\frac{1}{N_0}
\sum_{k_2}^{N_0} \sum_{j_1}^{N_*}
e^{-{\bf p}_{j_1} \cdot {\bf q}_{j_1,k_2} /\mathrm{i}\hbar} .
\end{eqnarray}
The mixed factor is obviously intensive
(i.e.\ independent of $N_0$ and of $N_*$)
in the thermodynamic limit,
since for each ground momentum state boson
there is a limited number of excited momentum state bosons
in the neighborhood that will give a non-zero weight
after averaging.

The singularly excited  mixed loop phase function
defined by the final equality here,
$\eta^{(l-1;1)}_{0*}({\bf q},{\bf p})$,
is identical to the mixed dimer phase function,
Eq.~(\ref{Eq:eta2MixPhase}).
Hence its average is also the same.
Importantly, it is independent of $l$,
which means that their total contribution is
\begin{eqnarray} \label{Eq:OmegaMix1}
-\beta \Omega_\mathrm{mix}^{(1)}
& = &
\sum_{l=2}^{N_0+1}
\left\langle \eta^{(l-1;1)}_{0*} /N_0
\right\rangle_{N_0,N_*,\mathrm{cl}}
\nonumber \\ & = &
N_0 \left\langle \eta^{(1;1)}_{0*} /N_0
\right\rangle_{N_0,N_*,\mathrm{cl}}
\nonumber \\ & = &
\frac{N_0N_*}{N}
\left\langle \tilde \eta^{(2)} /N
\right\rangle_{N,\mathrm{cl}}^\mathrm{corr} ,
\end{eqnarray}
where the final factor on the right hand side
is given in Eq.~(\ref{Eq:MixDimer}).
This is the total from all mixed symmetrization loops
that have a single excited momentum state boson.

One can write the grand potential as
$\Omega = \Omega_\mathrm{cl} + \Omega_* + \Omega_\mathrm{mix}$,
with the first two terms being given by
Eqs~(\ref{Eq:OmegaCl}) and (\ref{Eq:OmegaPure}), respectively.
The present $\Omega_\mathrm{mix}^{(1)}$
is the leading order contribution to the mixed term.
Its derivative at constant $N$ is
\begin{equation} \label{Eq:dWmix1}
\left(
\frac{\partial (-\beta \Omega_\mathrm{mix}^{(1)})}{\partial N_*} \right)_{N}
=
\frac{N  - 2 N_*}{N}
\left\langle \tilde\eta^{(2)}/N \right\rangle_{N,\mathrm{cl}}^\mathrm{corr}.
\end{equation}
The average is independent of $N_*$.
The prefactor is negative for $N_* > N/2$,
which is the case at high temperatures.
As mentioned above the average of the mixed dimer
must be negative for high temperatures.
These two facts means that at high temperatures
this derivative is positive,
which increases the occupation of the excited states
compared to the classical and pure terms alone.

\subsubsection{Numerical Results}

\begin{table}[tb]
\caption{ \label{Tab:tg2}
Liquid density, mixed dimer weight,
and optimum ground momentum state fraction
for saturated Lennard-Jones He$^4$ at various temperatures,
$T^* \equiv k_\mathrm{B} T /\varepsilon$.
From liquid drop Monte Carlo simulations with $N=5000$,
overall density of $\rho \sigma^3=0.2$,
using the central liquid volume of radius $ 10 \sigma$
for the averages.
}
\begin{center}
\begin{tabular}{c c c c c c}
\hline\noalign{\smallskip}
$ T^* $ & $\rho \sigma^3$ & $\rho \Lambda^3 $  &
$\displaystyle \frac{
\langle \tilde \eta^{(2)} \rangle_{N,\mathrm{cl}}^\mathrm{corr}
}{N} $ &
$\displaystyle \frac{\overline N_0}{N} $
& $\displaystyle \frac{\overline N_0}{N} $ \\
&&&& stable & unstable \\
\hline \rule{0cm}{0.4cm}%
1.00 & 0.67 & 0.81 & -0.60 & 0 & - \\
0.90 & 0.73 & 1.05 & -0.70 & 0 & - \\
0.80 & 0.79 & 1.34 & -0.79 & 0 & - \\
0.70 & 0.83 & 1.73 & -0.87 & 0 & - \\
0.65 & 0.86 & 1.99 & -0.90 & 0 & - \\
0.60 & 0.88 & 2.29 & -0.93 & 0.640 & 0.35 \\
0.55 & 0.90 & 2.67 & -0.96 & 0.754 & 0.35 \\
0.50 & 0.91 & 3.14 & -0.99 & 0.818 & 0.35 \\
0.45 & 0.97 & 3.93 & -1.04 & 0.872 & 0.45 \\
0.40 & 0.99 & 4.74 & -1.06 & 0.902 & 0.55 \\
\hline
\end{tabular} \\
\end{center}
\end{table}

Simulation results for the mixed dimer are shown in Table~\ref{Tab:tg2}.
The difference in the Lennard-Jones saturation density
compared to that in Table~\ref{Tab:dens}
is partly due to the larger system size and lower overall density here,
which means that the periodic boundary conditions
have less influence on the shape of the liquid phase.
The primary difference is that here the averages were
taken over a sphere of radius $10\sigma$ about the center of mass,
whereas in Table~\ref{Tab:dens} they were taken over the whole system,
of which approximately 20\% of the bosons were in the vapor phase
or interfacial region.
For the present larger system at $T^*=0.5$,
the radius of the interface at half density  is $\approx 10.6\sigma$.
The central density measured within $2\sigma$ of the center of mass is
$\rho\sigma^3=0.943(1)$,
whereas the density measured within $10\sigma$
is $\rho\sigma^3=0.9116(5)$.
For $T^* \alt 0.5$ the system was somewhat glassy,
with limited or no macroscopic diffusion of the particles.


It can be seen in Table~\ref{Tab:tg2}
that the mixed dimer weight is negative at all temperatures.
This means that the asymptotic correction dominates.
The weight increases in magnitude
with decreasing temperature,
although the dependence on temperature is rather weak.

The mixed dimer weight was combined
with the classical pure loop grand potential,
$\Omega = \Omega_\mathrm{cl} + \Omega_\mathrm{pure}
+ \Omega_\mathrm{mix}^{(1)}$,
with the first two terms being given by
Eqs~(\ref{Eq:OmegaCl}) and (\ref{Eq:OmegaPure}),
and the singly excited mixed grand potential by Eq.~(\ref{Eq:OmegaMix1}).
The pure loops used $l_\mathrm{max} = 4$ for $T^* \ge 0.7$
and $l_\mathrm{max} = 5$ for $T^* \le 0.65$.
The results were little different when $l_\mathrm{max}$ was increased by one.
The optimum fraction of bosons in the ground momentum state,
$\overline N_0/N$,
corresponds to the vanishing of the derivative of the grand potential,
Eq.~(\ref{Eq:dW/dN*}) plus Eq.~(\ref{Eq:dWmix1}).
It can be seen in Table~\ref{Tab:tg2}
that for $T^* \ge 0.65$ there is only one stable solution,
namely all the bosons in the system are in excited states.
For temperatures $T^* \le 0.6$ (6.13\,K),
there are two zeros for the derivative,
the higher fraction being the stable solution
(i.e.\ the minimum in the grand potential).
At $T^* = 0.6$ the stable fraction of ground momentum state bosons
is $\overline N_0/N = 0.640$,
which is a rather abrupt change from
$\overline N_0/N = 0$ at $T^*=0.65$.
It would be fair to call this ground momentum state condensation.
The transition more or less coincides with the passage of
$\overline N_0/N $ from greater than one half to less than one half,
as given by the  pure loops only  in Table~\ref{Tab:dens}.
The value $\overline N_0/N = 0.640$ obtained by including the mixed dimer
is greater than the value $\overline N_0/N = 0.536$ in Table~\ref{Tab:dens}
obtained with the pure loops only,
as is expected for a negative values of
$\langle \tilde \eta^{(2)} \rangle_{N,\mathrm{cl}}^\mathrm{corr}$.


In Table~\ref{Tab:tg2}
it can be seen that
$\langle \tilde \eta^{(2)} \rangle_{N,\mathrm{cl}}^\mathrm{corr}/N
> -1 $ for $T^* > 0.45$.
Obviously there is some uncertainty in where it first exceeds this bound
as it is the difference between two positive comparable quantities,
and it therefore has a larger relative error than either alone.
Minus one is significant because it marks the point,
at the dimer level of approximation,
$1 + (N_*/N) \langle \tilde \eta^{(2)}
\rangle_{N,\mathrm{cl}}^\mathrm{corr}/N$,
beyond which it is not possible to have
a fully excited system, $\overline N_*/N = 1$
(c.f.\ dressed bosons below).

In any case the present mixed loop calculations
provide a credible mechanism for ground momentum state condensation.
Compared to the unmixed results, Table~\ref{Tab:dens},
where the condensation grows continuously from zero at $\rho \Lambda^3 = 1$,
in the mixed case the condensation is rather sudden,
and it occurs at $\rho \Lambda^3 = 2.29$.
Since mixed loops are forbidden in the ideal gas
this effect is specific to interacting bosons.
It shows
that the growth of excited state position permutation loops
nucleate, or  are responsible for, ground momentum state condensation
(Attard 2021 section~5.6.1.7).

\subsection{Dressed  Bosons} \label{Sec:Dressed}

In the preceding subsection
the non-local nature of the ground momentum state
was exploited to incorporate an arbitrary number of
ground momentum state bosons into the mixed dimer permutation loop
to obtain the so-called singly excited grand potential
as the leading order contribution
from mixed ground and excited momentum state permutation loops.
In this subsection this idea is generalized to form mixed permutation
loops by concatenating permutation chains,
which consist of a ground momentum state boson as the head
and excited momentum state bosons as the tail.
The series of  such chains may be called
a dressed ground momentum state boson,
or dressed boson for short.
The theory based on them is like the ideal solution theory
of physical chemistry, with the dressed  ground momentum state bosons
being the solute that forms a dilute solution
in the fluid of excited  momentum state bosons.

Consider a $l$-chain,
with the ground momentum state boson at the head,
which we designate as position $l$,
and with $l-1$ excited momentum state bosons forming the tail.
A particular chain has symmetrization function
\begin{eqnarray}
\lefteqn{
\tilde \eta^{(l)}_{j_1,\ldots,j_l}
}  \\ \nonumber
& = &
e^{ -{\bf p}_{j_1} \cdot {\bf q}_{j_1,j_2} /\mathrm{i}\hbar}
e^{ -{\bf p}_{j_2} \cdot {\bf q}_{j_2,j_3} /\mathrm{i}\hbar}
\ldots
e^{ -{\bf p}_{j_{l-1}} \cdot {\bf q}_{j_{l-1},j_l} /\mathrm{i}\hbar} .
\end{eqnarray}

The sum of all possible dimer chains
is the same as the mixed dimer symmetrization function,
Eq.~(\ref{Eq:eta2MixPhase}),
\begin{eqnarray}
\eta^{(1;1)}_{0*}({\bf q},{\bf p})
& = &
\sum_{j}^{N_0}  \sum_{k}^{N_*}
 e^{-{\bf p}_{k} \cdot {\bf q}_{kj}/\mathrm{i}\hbar}
 \nonumber \\ & = &
\sum_{j}^{N_0}  \sum_{k}^{N_*}
\tilde \eta^{(2)}_{k,j}.
\end{eqnarray}
The average of this correcting for the asymptote is given as
Eq.~(\ref{Eq:MixDimer}).

The sum of all possible trimer chains
is the same as the mixed trimer symmetrization function
with one ground and two excited momentum state bosons,
Eq.~(\ref{Eq:eta(3;2)}),
\begin{eqnarray}
\eta^{(1;2)}_{0*}({\bf q},{\bf p})
& = &
\sum_{j}^{N_0}  \sum_{k,l}^{N_*}\!'\;
e^{-{\bf p}_{k} \cdot {\bf q}_{kl}/\mathrm{i}\hbar}
e^{-{\bf p}_{l} \cdot {\bf q}_{lj}/\mathrm{i}\hbar}
  \nonumber \\ & = &
\sum_{j}^{N_0}  \sum_{k,l}^{N_*}\!'\;
\tilde \eta^{(3)}_{k,l,j} .
\end{eqnarray}
The prime on the sum indicates that no two indeces may be equal.
The average of this correcting for the asymptote is given as
Eq.~(\ref{Eq:<eta(3;2)>}).

For a ground state boson $k$,
chains of a given length may be summed over all excited momentum state bosons,
\begin{eqnarray}
\tilde \eta^{(l)}_{k}({\bf q},{\bf p})
=
\sum_{j_1,\ldots,j_{l-1}}^{N_*}\hspace{-.3cm}'\hspace{.3cm}
\tilde \eta^{(l)}_{j_1,\ldots,j_{l-1},k}({\bf q},{\bf p}).
\end{eqnarray}
These in turn may be summed over all lengths
\begin{eqnarray}
\tilde \eta_{k}({\bf q},{\bf p})
=
\sum_{l=2}^{N_*+1}
\tilde \eta^{(l)}_{k} ({\bf q},{\bf p}).
\end{eqnarray}
The quantity
$\tilde \eta_{k}({\bf q},{\bf p})$
is the symmetrization weight of the dressed ground momentum state boson $k$
at this point in phase space.
We may average this weight,
either before or after summing over chain length,
and treat the dressed ground momentum state bosons as independent,
which is valid if they are dilute compared
to the excited momentum state bosons,
$N_0 (l_\mathrm{max}-1) \ll N_*$.

One can form $N_0!$ permutations of the dressed bosons.
But for this to work one needs to show
that the weight upon concatenation of two chains into a permutation loop
is the same as the product of the weights
of the two individual permutation chains.
To this end, consider an $l$-chain $j_1,j_2,\ldots, j_l$
and an $m$-chain, $k_1,k_2,\ldots, k_m$.
The symmetrization function for the permutation loop formed from them is
\begin{eqnarray}
\lefteqn{
\tilde \eta^{(l+m)}_{j^l,k^m}({\bf q},{\bf p})
} \nonumber \\
& = &
e^{ -{\bf p}_{j_1} \cdot {\bf q}_{j_1,j_2} /\mathrm{i}\hbar}
\ldots
e^{ -{\bf p}_{j_{l-1}} \cdot {\bf q}_{j_{l-1},j_l} /\mathrm{i}\hbar}
\nonumber \\ &&  \mbox{ } \times
e^{ -{\bf p}_{k_1} \cdot {\bf q}_{k_1,k_2} /\mathrm{i}\hbar}
\ldots
e^{ -{\bf p}_{k_{m-1}} \cdot {\bf q}_{k_{m-1},k_m} /\mathrm{i}\hbar}
\nonumber \\ & = &
\tilde \eta^{(l)}_{j^l}({\bf q},{\bf p})\,
\tilde \eta^{(m)}_{k^m}({\bf q},{\bf p}) .
\end{eqnarray}
The two transposition factors that link the chains,
$e^{ -{\bf p}_{j_{l}} \cdot {\bf q}_{j_{l},k_1} /\mathrm{i}\hbar}$
and
$e^{ -{\bf p}_{k_{m}} \cdot {\bf q}_{k_{m},j_1} /\mathrm{i}\hbar}$,
are unity because the head bosons are in the ground momentum state,
${\bf p}_{j_{l}}={\bf p}_{k_{m}}={\bf 0}$.
Hence the permutation loop function formed by concatenating
the two chains is just the product of the two chain functions,
with the restriction that no index can be the same.
Since the chains are compact,
and since it is assumed that $N_0 \ll N_*$ (dilute solution),
this restriction can be dropped with negligible error.
From this it follows that the $N_0!$ permutations of particular chains
form legitimate permutation loops,
and that all $N_0!$ permutations have the same total weight,
which is equal to the product of the weights of the individual chains.

The average weight of a dressed  ground momentum state boson is
\begin{eqnarray}
\left\langle
\tilde{\eta}({\bf q},{\bf p})
\right\rangle_{N_0,N_*,\mathrm{cl}}
& = &
\left\langle
\frac{1}{N_0} \sum_{k}^{N_0} \tilde \eta_{k}({\bf q},{\bf p})
\right\rangle_{N_0,N_*,\mathrm{cl}}
\nonumber  \\ & = &
\sum_{l=2}^{\infty}
\left\langle
\frac{1}{N_0} \sum_{k}^{N_0} \tilde \eta_{k}^{(l)}({\bf q},{\bf p})
\right\rangle_{N_0,N_*,\mathrm{cl}}
\nonumber  \\ & = &
\sum_{l=2}^{\infty}
\frac{N_*^{l-1}}{N^l} \left\langle
\tilde G^{(l)}
\right\rangle_{N,\mathrm{cl}}^\mathrm{corr}
\nonumber  \\ & = &
\sum_{l=2}^{\infty} \frac{N_*^{l-1}}{N^{l-1}} \tilde g^{(l)}.
\end{eqnarray}
The $l=2$ term is the mixed dimer, Eq.~(\ref{Eq:MixDimer}),
and the $l=3$ term is the mixed trimer, Eq.~(\ref{Eq:MixTrimer}).
The chain Gaussian is
\begin{eqnarray}
\tilde G^{(l)}({\bf q}^N)
& = &
\sum_{j_1,\ldots,j_l}^N\hspace{-.2cm}'\hspace{.1cm}
\prod_{k=1}^{l-1}
e^{-\pi q_{j_k,j_{k+1}} /\Lambda^2 } .
\end{eqnarray}
The prime indicates that no two indeces may be equal.

The  product of the dressed weights
of all the bosons  gives the mixed grand partition function,
\begin{equation}
\tilde \Xi
=
1 + \eta_\mathrm{mix}
=
\big[ 1 + \langle \tilde{\eta}({\bf q},{\bf p})
 \rangle_{N_0,N_*,\mathrm{cl}}  \big]^{N_0},
 \end{equation}
where the bare boson
(monomer chain), $\tilde \eta^{(1)}  = 1$,
appears explicitly.
The logarithm gives the mixed grand potential
\begin{equation}
-\beta \tilde\Omega
=
N_0 \ln
\big[ 1 + \langle \tilde{\eta}({\bf q},{\bf p})
 \rangle_{N_0,N_*,\mathrm{cl}}  \big]  .
\end{equation}
The total grand potential is
$\Omega = \Omega_\mathrm{cl} + \Omega_* + \tilde\Omega$.
Implicitly,
$N_0 = \overline N_0(z,V,T)$ and $N_* = \overline N_*(z,V,T)$,
although as a variational formulation
this is a second order effect.
The $N_0!$ permutations of the chains,
which is the pure ground momentum state symmetrization weight,
cancels with the same factor
in the denominator of the partition function.
The total weight of the ground momentum state bosons
is the product of their average dressed weight,
assuming that they do not overlap.

If one linearizes the expression for
the mixed grand potential one obtains
\begin{eqnarray}
-\beta \tilde\Omega
& = &
N_0
\sum_{l=2}^{\infty}
\left\langle
\frac{1}{N_0} \sum_{k}^{N_0} \tilde \eta_{k}^{(l)}({\bf q},{\bf p})
\right\rangle_{N_0,N_*,\mathrm{cl}}
\nonumber \\ & = &
N_0
\sum_{l=2}^{\infty} \frac{N_*^{l-1}}{N^{l-1}} \tilde g^{(l)}
\nonumber \\ & = &
-\beta \sum_{l=2}^{\infty}  \Omega_\mathrm{mix}^{(l-1)} .
\end{eqnarray}
The dimer term is just Eq.~(\ref{Eq:OmegaMix1})
and the trimer term is just Eq.~(\ref{Eq:OmegaMix2}).
This shows how the present dressed boson theory
is the non-linear generalization
of the singly and doubly excited loops
of section~\ref{Sec:MixDim} and appendix~\ref{Sec:MixTrim}.

The derivative of the non-linear mixed grand potential is
\begin{eqnarray}
\left(
\frac{\partial \big(\beta \tilde \Omega\big)}{\partial N_*}
\right)_N
& = &
-\ln  \left[ 1 +
\sum_{l=2}^\infty \frac{N_*^{l-1}}{N^{l-1}} \tilde g^{(l)} \right]
\nonumber \\ &  & \mbox{ }
+
\frac{ \displaystyle
\frac{N_0}{N}
\sum_{l=2}^\infty (l-1) \frac{N_*^{l-2}}{N^{l-2}}  \tilde g^{(l)}
}{ \displaystyle
1 +
\sum_{l=2}^\infty  \frac{N_*^{l-1}}{N^{l-1}}\tilde g^{(l)}} .
\end{eqnarray}
Add this to the unmixed result, Eq.~(\ref{Eq:dW/dN*}),
to obtain the fraction of excited momentum state bosons in the system.
One can also linearize this.

\begin{table}[tb]
\caption{ \label{Tab:tg3}
Chain weights,
$\tilde g^{(l)} =
\langle \tilde \eta^{(l)} \rangle_{N,\mathrm{cl}}^\mathrm{corr} /N$,
and optimum fraction of ground momentum state bosons
for saturated Lennard-Jones He$^4$ at various temperatures.
}
\begin{center}
\begin{tabular}{c c c c c c}
\hline\noalign{\smallskip}
$ T^* $ &
$ \displaystyle \frac{\langle \tilde{\eta}\!^{(2)}
\rangle_{N,\mathrm{cl}}^\mathrm{corr} }{N} $ &
$ \displaystyle \frac{ \langle \tilde{\eta}\!^{(3)}
\rangle_{N,\mathrm{cl}}^\mathrm{corr}  }{N} $ &
$ \displaystyle \frac{ \langle \tilde{\eta}\!^{(4)}
\rangle_{N,\mathrm{cl}}^\mathrm{corr}  }{N} $ &
$\displaystyle \frac{\overline N_0}{N} $
& $\displaystyle \frac{\overline N_0}{N} $ \\
&&&& linear & non-linear \\
\hline \rule{0cm}{0.4cm}%
1.00 & -0.599 & 0.357 & -0.213 & 0 & 0 \\
0.90 & -0.700 & 0.486 & -0.339 & 0 & 0 \\
0.80 & -0.790 & 0.613 & -0.485 & 0 & 0 \\
0.70 & -0.866 & 0.728 & -0.650 & 0 & 0 \\
0.65 & -0.904 & 0.782 & -0.754 & 0 & 0.527 \\
0.60 & -0.934 & 0.820 & -0.875 & 0.560 & 0.658 \\
0.55 & -0.963 & 0.852 & -1.058 & 0.677 & 0.745 \\
0.50 & -0.992 & 0.889 & -1.398 & 0.771 & 0.812 \\
0.45 & -1.038 & 1.089 & -2.496 & 0.845 & 0.868 \\
0.40 & -1.063 & 1.180 & -4.069 & 0.887 & 0.900 \\
\hline
\end{tabular} \\
\end{center}
\end{table}

Table~\ref{Tab:tg3} shows results for chains up to $l_\mathrm{max}=4$.
That the chain weights alternate in sign
poses convergence problems.
For $T^* \alt 0.55$ it is not clear that the chain series is converging
(although of course each term must be multiplied by $(N_*/N)^{l-1}$).
Retaining the first four terms in the series
(including the monomer term of unity)
gives a result for the optimum fraction of ground momentum state bosons
in broad agreement with those predicted retaining two terms only,
Table~\ref{Tab:tg2}.
The suppression of the occupation of the ground momentum state
for $T^* \ge 0.65$ is likewise clear in the two tables.
There is little difference between the linear and non-linear results,
which perhaps suggests that the linear theory is exact.
The location of the condensation,
$T^*=0.6$, $\rho \Lambda^3 = 2.29$,
is relatively insensitive to the level of the theory.

\begin{figure}[t!]
\centerline{ \resizebox{8cm}{!}{ \includegraphics*{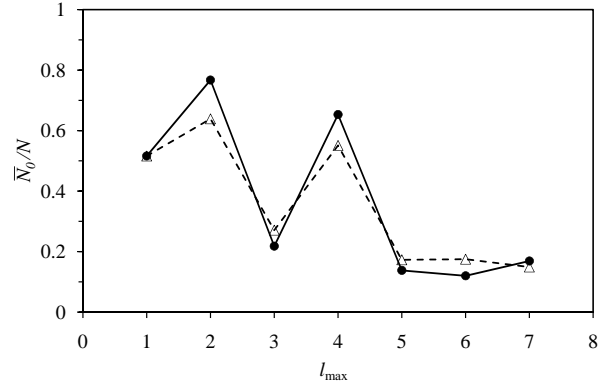} } }
\caption{\label{Fig:N0}
Fraction of ground momentum state bosons
as a function of maximum chain length
at $T^*=0.6$.
The circles are non-linear results
and the triangles are linear results,
with the lines being an eye-guide.
The unmixed excited momentum state loops used $l_\mathrm{max}=7$
in all cases.
The statistical error is on the order of 1\%.
}
\end{figure}

Figure~\ref{Fig:N0} shows the optimum fraction
of ground momentum state bosons
as a function of the maximum chain length
at $T^*=0.6$.
There is clearly an even/odd effect for $l_\mathrm{max}\le 5$,
but for larger values the fraction appears to have converged
to $\overline N_0/N \approx 0.15$.
Condensation appears to have occurred by $T^*=0.6$,
although for individual choices of $l_\mathrm{max}$
it can occur by $T^*=0.65$
(c.f.\ Table~\ref{Tab:tg3} for $l_\mathrm{max} = 4$, non-linear).

Similar behavior as a function of $l_\mathrm{max}$
occurred for $T^*=$ 0.62, 0.60, and 0.55.
Condensation had occurred for almost all $l_\mathrm{max}$,
linear and non-linear;
the two exceptions in the 42 cases calculated happened
at low $l_\mathrm{max}$.
Using  $l_\mathrm{max} = 7$,
the following ground momentum state bosons fractions were found:
For  $T^*=0.62$,
${\overline N_0}/{N} =$ 0.1480(3) (linear) and 0.1577(18) (non-linear).
For $T^*=0.6$    ,
${\overline N_0}/{N} =$ 0.1487(7) (linear) and 0.1695(23) (non-linear).
And for $T^*=0.55$
${\overline N_0}/{N} =$ 0.1405(4) (linear) and 0.1981(14) (non-linear).
Since the change in the condensed fraction
over these temperature intervals
is negligible compared to the change from zero at $T^*=0.65$,
one can conclude that the condensation transition is discontinuous.
The transition occurs somewhere in the range
$0.62 \le T_\mathrm{c}^* < 0.65$.

\subsection{Energy and Heat Capacity in the Condensed Regime}
\label{Sec:UCv}

In these results there is either no increase or else a slow increase
in the fraction of ground momentum state bosons
with decreasing temperature after the transition.
The slowness of the increase appears to be an artefact
that arises from assuming that the dressed bosons do not interfere
with each other.
The condition for non-interference, $N_0 (l_\mathrm{max}-1) \ll  N_*$,
holds barely if at all in the present case;
the left hand side is, say,
$ 5 \times 0.15 = 0.75$,
whereas the right hand side is $0.85$.
If interference were taken into account the contribution
from mixed loops would be reduced
(because many loops currently counted would be prohibited).
Since the mixed loop contribution suppresses
the number of ground momentum state bosons
compared to the pure loop prediction, Table~\ref{Tab:dens},
interference between the mixed loops would reduce their negative contribution
and hence increase the number of ground momentum state bosons
above that calculated in the preceding paragraph.
This means that the pure loop result becomes increasingly
accurate below the condensation transition,
as is its prediction that the ground momentum state is increasingly occupied
as the temperature is decreased, Table~\ref{Tab:dens}.

This discussion leads to the conclusion
that the mixed loop theory, linear or non-linear,
is probably accurate for predicting the location
of the condensation transition,
but that it underestimates the number of ground momentum state bosons,
and its rate of increase with decreasing temperature,
once condensation has occurred.
The pure loop theory is likely more accurate in the condensed regime.
In calculating the effect of condensation
on the energy and on the heat capacity,
our strategy will be to use the mixed loop theory to locate the transition,
and the pure loop theory to estimate
the number of ground momentum state bosons in the condensed regime.

The average energy is the sum of loop energies,
$\overline {\cal H} =\sum_{l=1}^\infty  \overline E^{(l)}$.
The averages are classical.
One can relate each average loop energy for a system
of $\overline  N_0$ ground momentum state bosons
and $\overline  N_* = N-\overline  N_0$ excited momentum state bosons,
$\overline E^{(l)}\!(\overline  N_0,\overline  N_*)$,
to one obtained in a classical system of $N$ bosons,
$\overline E^{(l)}\!(N)$, as follows.
The classical or monomer contribution
is the sum of the classical kinetic energy
and the potential energy,
\begin{equation}
\overline E^{(1)}\!(\overline  N_0,\overline  N_*)
=
\frac{3}{2} \overline  N_* k_\mathrm{B}T
+ \left\langle U \right\rangle_{N,V,T}^\mathrm{cl} .
\end{equation}
Obviously only excited momentum state bosons
contribute to the kinetic energy.
The potential energy does not distinguish
ground and excited momentum state bosons.
The dimer and higher loop energies are
\begin{equation}
\overline E^{(l)}\!(\overline  N_0,\overline  N_*)
=
\left( \frac{\overline  N_*}{N} \right)^{l}
\overline E^{(l)}\!(N),
\;\; l \ge 2.
\end{equation}
Above the condensation transition, $\overline  N_* = N$
and the two systems are the same.

\begin{figure}[t!]
\centerline{ \resizebox{8cm}{!}{ \includegraphics*{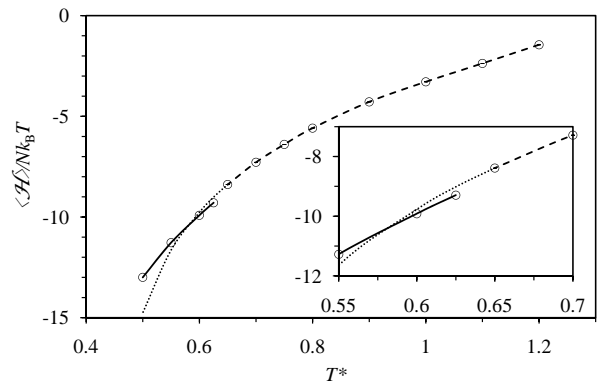} } }
\caption{\label{Fig:H}
Canonical Monte Carlo (homogeneous, $N=5,000$)
results  for the average energy for Lennard-Jones He$^4$
along the saturation curve  using $l_\mathrm{lmax}=5$.
The dashed and dotted curves use $N_*=N$,
whereas the solid curve uses $\overline N_0$
given by the pure loop theory with $l_\mathrm{max}=7$.
The error bars,
whose total length is four times the standard statistical error,
are less than the size of the symbols.
\textbf{Inset.} Focus on the transition.
}
\end{figure}

The average energy resulting from these formulae
is shown in Fig.~\ref{Fig:H}.
These data resulted from homogeneous canonical Monte Carlo
simulations along the saturated density curve.
The condensation transition was set at $T_\mathrm{c}^* = 0.625$,
as determined by the mixed loop theory described in the preceding subsection.
The fraction of ground momentum state bosons
below the condensation transition was
determined by the unmixed theory of section~\ref{Sec:NoMix} at the
heptamer level of approximation.
One can see in  Fig.~\ref{Fig:H}
that the condensed branch energy (solid curve) does not  coincide
with the non-condensed branch extrapolated to lower temperatures
(dotted curve). At the highest condensed temperature in the figure,
$T^* = 0.625$, the value of the extrapolated energy that neglects
the difference between ground and excited momentum state bosons is
$\beta \langle {\cal H} \rangle /N = -9.017(2)$, whereas the actual
condensed energy is $-9.295$. Hence the latent heat for the
Lennard-Jones condensation transition is
$ E_\mathrm{latent} = 0.3Nk_\mathrm{B}T$,
which is 3\% of the total energy.
This surprisingly small value results from three effects:
First, the kinetic energy, which is positive,
is lower in magnitude in the condensed regime.
Second, the loop energies, which are negative, are also lower in magnitude in
the condensed regime.
And third, the classical potential energy,
$\beta \overline U^{(1)} /N =  -9.958(1)$, which is 90\% of the total,
is unchanged.
Because at the transition the fraction of ground momentum
bosons is on the order of 50\% (according to the unmixed theory),
each individual change in energy is large relative to its
contribution to the total. But their relative contribution to the
total is only about 10\%, and the two changes partially cancel each other.
This leaves a residual latent heat of 3\% of the total energy.

The heat capacity, which is the temperature derivative of the energy,
can be similarly analyzed.
The monomer term is
\begin{equation}
C_\mathrm{V}^{(1)}(\overline  N_0,\overline  N_*)
=
\frac{3\overline N_*k_\mathrm{B}}{2}
+
\frac{3k_\mathrm{B}T}{2} \frac{\partial \overline  N_*}{\partial T}
+ C_\mathrm{V}^{\mathrm{cl,ex}}(N) .
\end{equation}
The dimer and higher loops contribute
\begin{eqnarray}
C_\mathrm{V}^{(l)}(\overline  N_0,\overline  N_*)
& = &
l \left( \frac{\overline  N_*}{N} \right)^{l-1}
\frac{\partial \overline  N_*}{N\partial T}
\overline E^{(l)}\!(N)
\nonumber \\ && \mbox{ }
+
\left( \frac{\overline  N_*}{N} \right)^{l}
C_\mathrm{V}^{(l)}(N) ,
\;\; l \ge 2.
\end{eqnarray}

The results are plotted in Fig.~\ref{Fig:CvLJ}.
The data are from homogeneous canonical Monte Carlo simulations
at the saturated density.
The difference with Fig.~\ref{Fig:Cv} is that those results
are $C_V(N)$, which assume that all $N$ bosons are excited,
whereas the results in  Fig.~\ref{Fig:CvLJ}
are $C_\mathrm{V}(\overline  N_0,\overline  N_*)$,
which take into account the number of ground momentum state bosons
given by the pure loop theory below the condensation transition
given by the mixed loop theory.
The qualitative and quantitative resemblance to the
measured $\lambda$-transition in liquid helium-4 is unmistakable.

The above evidence that the condensation transition for interacting bosons
is discontinuous is important.
Of course prior to condensation
the number of ground momentum state bosons is sub-extensive.
Any finite value for the fraction at the transition
indicates that the number of  ground momentum state bosons
has become extensive and it represents macroscopic condensation.
The present results show
that the system goes discontinuously from a sub-extensive (i.e.\ microscopic)
to an extensive (i.e.\ macroscopic) number of ground momentum state bosons
at the condensation transition.
This contrasts with the ideal gas,
where, in the thermodynamic limit,
the number of ground momentum state bosons
appears to grow continuously from zero
at the condensation transition
(c.f.\ the discussion following Table~\ref{Tab:dens}).

Although this observation provides a way of reconciling the experimental facts
that the superfluid and the $\lambda$-transition are coincident,
and that the superfluid transition is discontinuous,
it apparently contradicts the experimental measurements of helium-4,
which report the energy  as continuous (i.e.\ no latent heat)
and  the heat capacity as finite
at the $\lambda$-transition (Donnelly  and  Barenghi  1998),
which point is further discussed in the conclusion.
It is possible that the relatively small latent heat of 3\%,
as estimated here,
has been simply overlooked during temporal averaging
of dynamic heat capacity measurements.
This puzzle notwithstanding,
the present data suggest that superfluidity occurs
when the number of ground momentum
state bosons becomes extensive.
On the far side of the $\lambda$-transition
the heat capacity decreases
as the number of ground momentum state bosons increases.

\subsection{Dressed Bosons are Ideal}

The non-linear form for the weight of dressed ground momentum state bosons
has the appearance of a fugacity,
and so in the grand canonical system
one can make the replacement
$z^{N_0} \Rightarrow
\big(z [ 1 + \langle \tilde{\eta}
\rangle_{N_0, N_*,\mathrm{cl}}  ]\big)^{N_0}$.
One then has a form of ideal solution theory,
with the  dressed ground momentum state bosons
acting as dilute solutes in a fluid of excited momentum state bosons.
In this case the ground momentum state contribution to the grand potential,
including the mixed loops with excited momentum state bosons,
is that of an effective ideal gas,
\begin{equation}
-\beta \tilde\Omega(z,V,T)
\approx
- \ln \big\{ 1 -
z [ 1 + \langle \tilde{\eta}
\rangle_{\overline N_0,\overline N_*,\mathrm{cl}}  ]\big\}.
\end{equation}
Compare this with the first term in Eq.~(\ref{Eq:OmegaIdGas}).
This shows formally how interactions between bosons
modify the ground momentum state contribution
to the ideal gas  grand potential.
To this should be added the classical
and the pure excited state contributions,
$\Omega_\mathrm{cl} + \Omega_*$.

If $-1 < \langle \tilde{\eta}
\rangle_{\overline N_0,\overline N_*,\mathrm{cl}} < 0$,
which is likely the case on the high temperature side of the transition,
then the effective fugacity is less than the actual fugacity,
and the number of ground momentum state bosons
would be less than predicted by neglecting mixed loops.
For  $\langle \tilde{\eta}
\rangle_{\overline N_0,\overline N_*,\mathrm{cl}} < -1$,
the effective  fugacity would be negative,
which would be problematic.
Each term in the series of chain weights in Table~\ref{Tab:tg3}
has to be multiplied by $(N_*/N)^{l-1}$,
and so one cannot say \emph{a priori}
that the total is less than $-1$
without taking this into account.
On the other hand, what one \emph{can} see in Table~\ref{Tab:tg3}
is that for the fully excited system, $N_*/N=1$,
the sum  $ l \in [2 , 4]$ is less than $-1$ for $T^* < 0.6$.
This is thus the spinodal limit for full excitation,
at least at this level of approximation.
(The sum   $ l \in [2 , 7]$  at $T^*=0.6$ is $1.4 \pm .1$,
which says that the spinodal limit has not yet been reached
at this temperature
at the $l_\mathrm{max}=7$ level of approximation.)

%
\section{Conclusion}
\setcounter{equation}{0} \setcounter{subsubsection}{0}
%

\subsection{The $\lambda$-Transition }

\emph{Momentum states are better than energy states
for representing interacting particles.}
For the ideal gas the two representations are essentially equivalent,
but in going beyond the ideal gas one actually has a choice.
Perhaps the ubiquity of the energy representation in quantum mechanics
is the reason that  workers have not hitherto averted to
the advantages offered by the alternative.
In general for quantum particles that interact with non-zero
pair and many-body potentials,
momentum states have the advantage
because the eigenvalues and eigenfunctions are explicit and analytic.
Consequently the mathematical analysis is generally simple
and transparent,
the computer algorithms can be very efficient,
and the physical interpretation quite straightforward.
Non-commutativity effects that occur for momentum states
become unimportant on the long length scales
that dominate  the $\lambda$-transition and superfluidity.
For  Bose-Einstein condensation specifically,
the form of the symmetrized bosonic wave function
that is ultimately responsible for the phenomenon
is particularly simple to derive and to manipulate
because, unlike energy states, momentum state are single particle states.
For the phenomenon of superfluidity,
momentum states are vector states
whereas energy states are scalar states.
Only the former couple directly to hydrodynamic flow,
which is likewise a vector quantity.
Finally it is quite easy to demonstrate explicitly the non-local nature
of the momentum ground state,
and it is uniquely this  that is responsible both for  Bose-Einstein
condensation and for superfluidity.

\emph{Interactions between bosons significantly contribute
to the magnitude and growth rate  of the heat capacity
approaching the $\lambda$-transition from the high temperature side
in liquid helium-4.}
The heat capacity predicted by London's (1938) ideal gas treatment
is substantially smaller than the measured values,
and the predicted growth rates qualtiatively differ from
the measured growth and decay
on either side of the $\lambda$-transition.
The present computer simulation results
for bosons interacting with the Lennard-Jones pair potential
are much closer to the measured values for the heat capacity
in magnitude and form,
and they predict a steep growth approaching the $\lambda$-transition.
The mechanism for the growth
on the high temperature side of the $\lambda$-transition
is the growth in number, size, and extent
of excited momentum state position permutation loops
(Attard 2021 chapter~5).
The specific origin of this growth
is the peak at contact in the many-particle correlation function
in the interacting system.
Such an adsorption excess does not occur in the ideal gas.

\emph{The decrease in the heat capacity on the low temperature side
of the $\lambda$-transition
is due to the increasing occupancy of the momentum ground state.}
Although this effect is also qualitatively captured by the ideal gas,
interactions make a difference because
they allow mixed ground and excited momentum state permutation loops.
Mixed permutation loops are forbidden in the ideal gas.
Mixed loops suppress the occupation of the ground momentum state
above the $\lambda$-transition,
and determine quantitatively the fraction of bosons
in the ground momentum state below it.
Condensation due to the coupling provided by mixed permutation loops
terminates the growth of the heat capacity.
This defines the $\lambda$-transition,
and forces it to coincide with the superfluid transition.
The latter is a manifestation
of Bose-Einstein condensation into the momentum ground state.

\emph{Mixed permutation loops cause
discontinuous ground momentum state condensation
at the $\lambda$-transition,
which gives a discontinuous transition to superfluidity,
and a latent heat for the transition,
which causes a divergence in the heat capacity.}
In the ideal gas the condensation is continuous,
there is no latent heat,
and the heat capacity does not diverge.
The experimental evidence currently shows no latent heat
or divergence in the heat capacity
(Donnelly  and  Barenghi  1998).
The present results show a discontinuity in the average energy
at the condensation transition, Fig.~\ref{Fig:H},
that is about 3\% of the total energy.
Perhaps this is beyond the resolution of experimental measurements,
or perhaps it is washed out by the time averaging
implicit in dynamic measurement procedures.
Establishing an upper bound for any measured latent heat
would provide a possible method to falsify the present theory
or the approximations that it invokes.

\subsection{Superfluid Flow}

The present theory is based upon momentum states,
whereas conventional approaches to  superfluidity
are based upon energy states.
As pointed out following Eq.~(\ref{Eq:N0!}),
all of the ground momentum state bosons in the system
participate in the pure permutation loops.
This is an example of quantum non-locality.
One can identify three specific consequences for superfluidity.
First, there can be no spatial correlation of bosons
in the ground momentum state,
and therefore there can be no spatial correlation
of ground state momentum.
This means that when an external field or pressure gradient
raises the ground state to be one with non-zero momentum,
then the consequent flow can only be plug flow.
In classical hydrodynamics this is known as inviscid flow.

Second,
when bosons condense into the momentum ground state,
collisions with the walls or with excited state momentum bosons
have an entropic component that involves the state as a whole.
In general an individual collision that excites a single boson from the
ground to an excited momentum state \emph{reduces} the permutation entropy
of the system.
Such collisions are suppressed
and only collisions that excite \emph{every} ground momentum state boson
increase the entropy (Attard 2021 Eq.~(5.69)).
This is the reason that ground momentum state bosons flow without viscosity
and without collisions with excited momentum state bosons.

\begin{figure}[t!]
\centerline{ \resizebox{8cm}{!}{ \includegraphics*{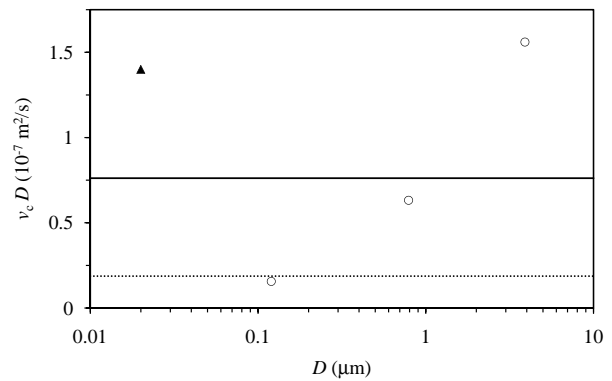} } }
\caption{\label{Fig:vc}
Critical velocity times the pore diameter
for superfluid flow
of helium through a cylindrical pore of diameter $D$.
The circles (Pathria 1972 section~10.8)
and triangle (Allum \emph{et al.}\ 1977)
are measured values,
the solid line is the present prediction
$v_r = 2 j_{01} \hbar/mD$,
and the dotted line is the vortex result of Kawatra and Pathria (1966)
$v_\mathrm{c} = 1.18 \hbar/mD$.
}
\end{figure}

And third, there is momentum gap due to the size of the system;
$\Delta_p = 2 \pi \hbar /L$ in rectangular geometry,
and $\Delta_{p,\perp} = 2 j_{01} \hbar /D$
in cylindrical geometry (Blinder 2011),
where $j_{01} = 2.4$ is the first zero of the zeroth order Bessel function.
This, combined with the second point,
means that the critical velocity for superfluid flow
is equal to the velocity of the first excited transverse state.
Although it has previously been known that the critical velocity
is on the order of $\hbar/mD$ (Fetter and Foot 2012),
the present theory gives a quantitative value,
a mathematical derivation, and a molecular justification for it
(Attard 2021).

This predicted critical velocity is tested in Fig.~\ref{Fig:vc}.
Over some three orders of magnitude in pore diameter
it remains within a factor of three of the measured data.
The Landau stability criterion for superfluid flow
gives a critical velocity that is several orders of magnitude larger
than the measured values (Batrouni \emph{et al.} 2004).
The vortex prediction of Kawatra and Pathria (1966),
which implements Feynman's (1954) suggestion
that the rotons invented by Landau (1941)
in his theory of superfluidity were in fact quantized vortices,
is also compared in the figure.
Although limited,
the data in Fig.~\ref{Fig:vc} lies closer to
the present theory than the vortex/roton theory.
More persuasive is Occam's apothegm:
of two explanations that equally describe the data,
the simpler is to be preferred.

\section*{References}

\begin{list}{}{\itemindent=-0.5cm \parsep=.5mm \itemsep=.5mm}

\item 
Abramowitz M and Stegun IA 1970
\emph{Handbook of Mathematical Functions}
(Dover: New York) 

\item 
Allum DR McClintock PVE and Phillips A 1977
The breakdown of superfluidity in liquid He:
an experimental test of Landau's theory
\emph{Phil.\ Trans.\ R.\ Soc.\ A} {\bf 284} 179

\item 
Attard P 2018
Quantum Statistical Mechanics in Classical Phase Space. Expressions for
  the Multi-Particle Density, the Average Energy, and the Virial Pressure
arXiv:1811.00730

\item 
Attard P 2021
\emph{Quantum Statistical Mechanics  in Classical Phase Space},
(IOP Publishing: Bristol).

\item 
Balibar S 2014
Superfluidity: how quantum mechanics became visible
pages 93--117
in
\emph{History of Artificial Cold, Scientific, Technological
and Cultural Issues}
(Gavroglu K editor)
(Dordrecht: Springer)

\item 
Balibar S 2017
Laszlo Tisza and the two-fluid model of superfluidity
\emph{C.\ R.\ Physique} {\bf 18}  586

\item
Batrouni GG, Ramstad T  and Hansen A 2004
Free-energy landscape and the critical
velocity of superfluid films
\emph{Phil.\ Trans.\ R.\ Soc.\ Lond.}\ A {\bf 362} 1595

\item 
Blinder SM 2011
Quantum-mechanical particle in a cylinder http://%
demonstrations.wolfram.com\\/QuantumMechanicalParticleInACylinder
\\/WolframDemonstrationsProject

\item
Donnelly RJ 1995
The discovery of superfluidity
Physics Today   {\bf 48} 30

\item
Donnelly RJ 2009
The two-fluid theory and second sound in liquid helium
Physics Today   {\bf 62} 34

\item
Donnelly RJ and  Barenghi CF 1998
The observed properties of liquid Helium at the saturated vapor pressure
\emph{J.\ Phys.\ Chem.\ Ref.\ Data} {\bf 27} 1217

\item
Fetter AL and Foot CJ 2012
Bose Gas: Theory and Experiment,
chapter 2 in Ultracold Bosonic and Fermionic Gases
Edited by Levin K,  Fetter AL, and Stamper-Kurn DM
(Elsevier:Amsterdam)

\item 
Feynman RP 1954
Atomic theory of the two-fluid model of liquid helium
\emph{Phys.\ Rev.}\ {\bf 94} 262

\item  
Kawatra MP and Pathria RK 1966
Quantized vortices in an imperfect
Bose gas and the breakdown of superfluidity in liquid helium II
\emph{Phys.\ Rev.}\ {\bf 151}  132

\item 
Landau LD 1941
Two-fluid model of liquid helium II
\emph{ J.\ Phys.\ USSR} {\bf 5} 71.
Also \emph{Phys.\ Rev.}\ {\bf 60} 356

\item 
London F 1938
The $\lambda$-phenomenon of liquid helium and the Bose-Einstein degeneracy
\emph{Nature} {\bf 141} 643

\item
Merzbacher E 1970
\emph{Quantum Mechanics} 2nd edn
(New York: Wiley)

\item 
Messiah A 1961 \emph{Quantum Mechanics}
(Vol 1 and 2) (Amsterdam: North-Holland)

\item 
Pathria RK 1972
\emph{Statistical Mechanics}
(Oxford: Pergamon Press)

\item
van Sciver  SW 2012
\emph{Helium Cryogenics}
2nd edn (New York: Springer)

\item 
Tisza L 1938
Transport phenomena in helium II
\emph{Nature} {\bf 141} 913

\end{list}

\comment{


} 

\appendix

%
\section{Mixed Trimer} \label{Sec:MixTrim}
\setcounter{equation}{0} \setcounter{subsubsection}{0}
\renewcommand{\theequation}{\Alph{section}.\arabic{equation}}
%

Following on from the mixed dimer of section~\ref{Sec:MixDim},
which is necessarily singly excited,
here is analyzed  the mixed trimer
with one ground and two excited momentum state bosons,
\begin{equation} \label{Eq:eta(3;2)}
\eta^{(1;2)}_{0*}({\bf q},{\bf p})
=
\sum_{j}^{N_0}  \sum_{k,l}^{N_*}\!'\;
 e^{-{\bf p}_{k} \cdot {\bf q}_{jl}/\mathrm{i}\hbar}
 e^{-{\bf p}_{l} \cdot {\bf q}_{lk}/\mathrm{i}\hbar} .
\end{equation}
The summand is just a trimer chain, section~\ref{Sec:Dressed}.
As for the dimer,
this should be divided by $N_0$ to account
for the fact that there are now $N_0-1$
ground momentum state bosons available for pure loops;
with this prefactor we can leave  $\eta_0=N_0!$ unchanged.
The prime on the sum over excited bosons indicates
that no two may be the same.
But  since the order matters each pair must be counted twice.
There are $N_0 N_*(N_*-1)$ such trimer loops,
most of which will contribute nothing after averaging.

The canonical average,
in the discrete momentum case is
\begin{eqnarray}
\lefteqn{
\left\langle \eta^{(1;2)}_{0*} /N_0 \right\rangle_{N_0,N_*,\mathrm{cl}}
}  \\
& = &
\frac{\Lambda^{3N_*}}{V^{N_*}}
\prod_j^{N_0} \delta_{{\bf p}_j,{\bf 0}}
\prod_{k}^{N_*} \sum_{{\bf p}_k}\!^{(p_k>0)} e^{-\beta p_k^2/2m }
 \nonumber \\ & & \mbox{ } \times
\left\langle
\frac{1}{N_0}
\sum_{j}^{N_0}  \sum_{k,l}^{N_*}\!'
 e^{-{\bf p}_{k} \cdot {\bf q}_{jl}/\mathrm{i}\hbar}
 e^{-{\bf p}_{l} \cdot {\bf q}_{lk}/\mathrm{i}\hbar}
 \right\rangle_{N,\mathrm{cl}}
 \nonumber \\ & = &
\frac{\Lambda^{3N_*}}{V^{N_*}}
\frac{N_*(N_*-1)}{N(N-1)(N-2)}
\nonumber \\ & & \mbox{ } \times
\prod_j^{N_0} \delta_{{\bf p}_j,{\bf 0}}
\prod_k^{N_*} \sum_{{\bf p}_k}\!^{(p_k>0)} e^{-\beta p_k^2/2m }
\int_V \mathrm{d}{\bf q}_1\,\mathrm{d}{\bf q}_2\,\mathrm{d}{\bf q}_3\;
 \nonumber \\ & & \mbox{ } \times
\rho^{(3)}_N({\bf q}_1,{\bf q}_2,{\bf q}_3)
 e^{-{\bf p}_{1} \cdot {\bf q}_{12}/\mathrm{i}\hbar}
 e^{-{\bf p}_{2} \cdot {\bf q}_{23}/\mathrm{i}\hbar} .\nonumber
\end{eqnarray}
In the final equality bosons 1 and 2 are in an excited momentum state
and boson 3 is in the momentum ground state,
although this makes no difference to the configuration integral.

The connected part of the triplet density is given by
\begin{eqnarray}
\lefteqn{
\rho^{3} h^{(3)}_N({\bf q}_1,{\bf q}_2,{\bf q}_3)
} \nonumber \\
& = &
\rho^{(3)}_N({\bf q}_1,{\bf q}_2,{\bf q}_3)
- \rho^3  h^{(2)}_N(q_{12})
- \rho^3 h^{(2)}_N(q_{23})
 \nonumber \\ & & \mbox{ }
- \rho^3 h^{(2)}_N(q_{31})
- 1 .
\end{eqnarray}
This vanishes if any particle is taken far from the rest.
Again we assume a homogeneous system.

The asymptotic contribution to the configurational integral is
\begin{eqnarray}
\lefteqn{
\rho^3
\int_V \mathrm{d}{\bf q}^3\;
\left\{
h^{(2)}_N(q_{12}) + h^{(2)}_N(q_{23}) + h^{(2)}_N(q_{31}) + 1
\right\}
 } \nonumber \\ && \mbox{  }
 \hspace{.7cm} \times
 e^{-{\bf p}_{1} \cdot {\bf q}_{12}/\mathrm{i}\hbar}
 e^{-{\bf p}_{2} \cdot {\bf q}_{23}/\mathrm{i}\hbar}
\nonumber \\
& = &
\rho^3 V \delta_{{\bf p}_2,{\bf 0}}
\int_V \mathrm{d}{\bf q}_{12}\; h^{(2)}_N(q_{12})
\nonumber \\ && \mbox{  }
+
\rho^3 V \delta_{{\bf p}_1,{\bf 0}}
\int_V \mathrm{d}{\bf q}_{23}\; h^{(2)}_N(q_{23})
\nonumber \\ & & \mbox{ }
+ \rho^3 V^2 \delta_{{\bf p}_1,{\bf p}_2}
\int_V \mathrm{d}{\bf q}_{13}\;
h^{(2)}_N(q_{31})
 e^{-{\bf p}_{1} \cdot {\bf q}_{13}/\mathrm{i}\hbar}
 \nonumber \\ && \mbox{  }
+
\rho^3 V \delta_{{\bf p}_1,{\bf 0}} \delta_{{\bf p}_2,{\bf 0}}
\nonumber \\ & = &
\rho^3 V^2 \delta_{{\bf p}_1,{\bf p}_2}
\int_V \mathrm{d}{\bf q}_{13}\;
h^{(2)}_N(q_{31})
 e^{-{\bf p}_{1} \cdot {\bf q}_{13}/\mathrm{i}\hbar} .
\end{eqnarray}
Since $p_1 > 0$ and $p_2 > 0$,
all the other terms vanish.

Separating out the surviving part of the asymptote
one obtains
\begin{eqnarray} \label{Eq:<eta(3;2)>}
\lefteqn{
\left\langle \eta^{(1;2)}_{0*} /N_0 \right\rangle_{N_0,N_*,\mathrm{cl}}
} \nonumber \\
& = &
\frac{\Lambda^{3N_*}}{V^{N_*}}
\frac{N_*(N_*-1)}{N(N-1)(N-2)}
\left\{
\Delta_p^{-3N_*}
\int \mathrm{d}{\bf p}^{N_*}\; e^{-\beta {\cal K}({\bf p}^{N_*}) }
\right. \nonumber \\ && \left. \mbox{ } \times
\int \mathrm{d}{\bf q}^3\,
\rho^3 h^{(3)}_N({\bf q}_1,{\bf q}_2,{\bf q}_3)
 e^{-{\bf p}_{1} \cdot {\bf q}_{12}/\mathrm{i}\hbar}
 e^{-{\bf p}_{2} \cdot {\bf q}_{23}/\mathrm{i}\hbar}
\right. \nonumber \\ && \left. \mbox{ }
 +
\Delta_p^{-3N_*}
\int \mathrm{d}{\bf p}^{N_*}\; e^{-\beta {\cal K}({\bf p}^{N_*}) }
\delta_{{\bf p}_1,{\bf p}_2}
\right. \nonumber \\ && \left. \mbox{ } \times
\rho^3 V^2
\int_V \mathrm{d}{\bf q}_{13}\;
h^{(2)}_N(q_{31})
e^{-{\bf p}_{1} \cdot {\bf q}_{13}/\mathrm{i}\hbar}
\right\}
 \nonumber \\ & = &
\frac{N_*(N_*-1)}{N(N-1)(N-2)}
\left\{
\rho^3\int \mathrm{d}{\bf q}_1\,\mathrm{d}{\bf q}_2\,\mathrm{d}{\bf q}_3\,
\right. \nonumber \\ && \left. \mbox{ } \times
 h^{(3)}_N({\bf q}_1,{\bf q}_2,{\bf q}_3)
 e^{-\pi q_{12}^2 /\Lambda^2}
 e^{-\pi q_{23}^2 /\Lambda^2}
\right. \nonumber \\ && \left. \mbox{ }
+
2^{-3/2} \rho^3 V \Lambda^{3}
\int_V \mathrm{d}{\bf q}_{13}\;
h^{(2)}_N(q_{13})  e^{-\pi q_{13}^2 /2\Lambda^2}
\right\}.
\end{eqnarray}
Both terms in the braces scale with volume
and hence this mixed trimer term is intensive.
The second term is likely negative,
but the first integrand has positive contributions.
This may be rewritten as classical canonical averages
\begin{eqnarray} \label{Eq:MixTrimer}
\lefteqn{
\left\langle \eta^{(1;2)}_{0*} /N_0 \right\rangle_{N_0,N_*,\mathrm{cl}}
} \nonumber \\
& =&
\frac{N_*^2}{N^2} \left\{
\left< \frac{1}{N} \sum_{j,k,l}^N\!'  e^{-\pi q_{jk}^2 /\Lambda^2}
 e^{-\pi q_{kl}^2 /\Lambda^2} \right>_{N,\mathrm{cl}}
\right. \nonumber \\ & & \mbox{ } \left.
- 2 \rho \Lambda^3
\left< \frac{1}{N} \sum_{j,k}^N\!'  e^{-\pi q_{jk}^2 /\Lambda^2}
\right>_{N,\mathrm{cl}}
+  \rho^2 \Lambda^6
\right\}
\nonumber \\ & =&
\frac{N_*^2}{N^2}
\left< \frac{1}{N} \sum_{j,k,l}^N\!'
\left[  e^{-\pi q_{jk}^2 /\Lambda^2} - \frac{\rho \Lambda^3}{N}  \right]
\right. \nonumber \\ && \left. \mbox{ } \times
\left[  e^{-\pi q_{kl}^2 /\Lambda^2} - \frac{\rho \Lambda^3}{N}  \right]
\right>_{N,\mathrm{cl}}
\nonumber \\ & \equiv &
\frac{N_*^2}{N^2}
\left\langle \tilde \eta^{(3)}  /N
\right\rangle_{N,\mathrm{cl}}^\mathrm{corr} .
\end{eqnarray}
Comparing the penultimate equality with Eq.~(\ref{Eq:MixDimer}),
the general rule can be inferred.

As for the mixed dimer, section~\ref{Sec:MixDim},
one can replace the single ground state boson
by the series of all the ground momentum state bosons,
each of the $N_0$ terms of which have the same value,
section~\ref{Sec:SingExLoop}.
Their total contribution is
\begin{eqnarray} \label{Eq:OmegaMix2}
-\beta \Omega_\mathrm{mix}^{(2)}
& = &
\sum_{l=3}^{N_0+2}
\left\langle \eta^{(l-2;2)}_{0*} /N_0
\right\rangle_{N_0,N_*,\mathrm{cl}}
\nonumber \\ & = &
N_0 \left\langle \eta^{(1;2)}_{0*} /N_0
\right\rangle_{N_0,N_*,\mathrm{cl}}
\nonumber \\ & = &
\frac{N_0 N_*^2}{N^2}
\left\langle \tilde \eta^{(3)} /N
\right\rangle_{N,\mathrm{cl}}^\mathrm{corr} .
\end{eqnarray}
The prefactor derivative  changes sign at $N_*/N = 2/3$.

%
\section{Corrigible Chains}
\setcounter{equation}{0} \setcounter{subsubsection}{0}
\renewcommand{\theequation}{\Alph{section}.\arabic{equation}}
%

The chain weights are corrected canonical averages,
\begin{equation}
\left\langle
\tilde{\eta}\!^{(l)} /N_0
\right\rangle_{N_0,N_*,\mathrm{cl}}
=
\frac{N_*^{l-1}}{N^{l-1}}
\left\langle
\tilde{\eta}\!^{(l)} /N
\right\rangle_{N,\mathrm{cl}}^\mathrm{corr} .
\end{equation}
Define the intensive uncorrected $l$-chain weight
\begin{equation}
\tilde g^{(l)} =
\frac{1}{N} \left\langle \sum_{j_1,\ldots,j_l}^N \!\!'\;
\prod_{k=1}^{l-1}  e^{-\pi q_{j_k,j_{k+1}}^2 /\Lambda^2}
\right\rangle_{N,\mathrm{cl}} .
\end{equation}
The prime on the summation
indicates that no two indeces may be the same.

This uncorrected chain weight
is the simplest to implement computationally.
Essentially this is because the head bosons
come from the liquid volume,
$N_\mathrm{liq}$,
and this is different to the number of neighbors,
${\cal N}_{j_1}$,
from which the tail of the chain is chosen.
The uncorrected weight is intensive and is insensitive to ${\cal N}_{j_1}$.

The relations between the first several
corrected and uncorrected chain weights are as follows.
(Here $N=N_\mathrm{liq}$ and $\rho = N_\mathrm{liq}/V_\mathrm{liq}$.
The product of the averages is used in the following expressions.)
The dimer is
\begin{eqnarray}
\left\langle
\tilde{\eta}\!^{(2)} /N
\right\rangle_{N,\mathrm{cl}}^\mathrm{corr}
& = &
\left\langle
\frac{1}{N}
\sum_{j,k}^N\!'
\left[  e^{-\pi q_{jk}^2 /\Lambda^2}
- \frac{\rho \Lambda^3}{N}  \right]
\right\rangle_{N,\mathrm{cl}}
\nonumber \\ & =&
\tilde g^{(2)} - \rho \Lambda^3.
\end{eqnarray}
The trimer is
\begin{eqnarray}
\left\langle
\tilde{\eta}\!^{(3)} /N
\right\rangle_{N,\mathrm{cl}}^\mathrm{corr}
& = &
\left< \frac{1}{N} \sum_{j,k,l}^N\!'
\left[  e^{-\pi q_{jk}^2 /\Lambda^2} - \frac{\rho \Lambda^3}{N}  \right]
\right. \nonumber \\ && \left. \mbox{ } \times
\left[  e^{-\pi q_{kl}^2 /\Lambda^2} - \frac{\rho \Lambda^3}{N}  \right]
\right>_{N,\mathrm{cl}}
\nonumber \\ & =&
\tilde g^{(3)}
- 2 \rho \Lambda^3 \tilde g^{(2)}
+ (\rho \Lambda^3)^2  .
\end{eqnarray}
The tetramer is
\begin{eqnarray}
\lefteqn{
\left\langle
\tilde{\eta}\!^{(4)} /N
\right\rangle_{N,\mathrm{cl}}^\mathrm{corr}
}  \\
& = &
\left<
\frac{1}{N} \sum_{j,k,l,m}^N\!'
\left[  e^{-\pi q_{jk}^2 /\Lambda^2} - \frac{\rho \Lambda^3}{N}  \right]
\right. \nonumber \\ && \left. \mbox{ } \times
\left[  e^{-\pi q_{kl}^2 /\Lambda^2} - \frac{\rho \Lambda^3}{N}  \right]
\left[  e^{-\pi q_{lm}^2 /\Lambda^2} - \frac{\rho \Lambda^3}{N}  \right]
\right>_{N,\mathrm{cl}}
\nonumber \\ & =&
\tilde g^{(4)}
- 2\rho \Lambda^3 \tilde g^{(3)}
- \rho \Lambda^3  \big(\tilde g^{(2)}\big)^2
+ 3 \rho^2 \Lambda^6  \tilde g^{(2)}
- \rho^3 \Lambda^9   .\nonumber
\end{eqnarray}
The pentamer is
\begin{eqnarray}
\lefteqn{
\left\langle
\tilde{\eta}\!^{(5)} /N
\right\rangle_{N,\mathrm{cl}}^\mathrm{corr}
} \nonumber \\
& = &
\tilde g^{(5)}
- 2\rho \Lambda^3 \tilde g^{(4)}
- 2\rho \Lambda^3\tilde g^{(3)} \tilde g^{(2)}
+  3 (\rho \Lambda^3)^2  \tilde g^{(3)}
\nonumber \\ && \mbox{ }
+ 3 (\rho \Lambda^3)^2  \big(\tilde g^{(2)}\big)^2
- 4 (\rho \Lambda^3)^3  \tilde g^{(2)}
+ (\rho \Lambda^3)^4  .
\end{eqnarray}
The hexamer is
\begin{eqnarray}
\lefteqn{
\left\langle
\tilde{\eta}\!^{(6)} /N
\right\rangle_{N,\mathrm{cl}}^\mathrm{corr}
} \nonumber \\
& = &
\tilde g^{(6)}
- 2\rho \Lambda^3 \tilde g^{(5)}
- 2\rho \Lambda^3 \tilde g^{(4)} \tilde g^{(2)}
- \rho \Lambda^3  \big(\tilde g^{(3)}\big)^2
 \nonumber \\ &&  \mbox{ }
+  2 (\rho \Lambda^3)^2  \tilde g^{(4)}
+ 7 (\rho \Lambda^3)^2  \tilde g^{(3)} \tilde g^{(2)}
+ (\rho \Lambda^3)^2   \big(\tilde g^{(2)}\big)^3
 \nonumber \\ &&  \mbox{ }
- 4 (\rho \Lambda^3)^3 \tilde g^{(3)}
- 6 (\rho \Lambda^3)^3  \big(\tilde g^{(2)}\big)^2
 \nonumber \\ &&  \mbox{ }
+ 5 (\rho \Lambda^3)^4  \tilde g^{(2)}
- (\rho \Lambda^3)^5  .
\end{eqnarray}
And the heptamer is
\begin{eqnarray}
\lefteqn{
\left\langle
\tilde{\eta}\!^{(7)} /N
\right\rangle_{N,\mathrm{cl}}^\mathrm{corr}
} \nonumber \\
& = &
\tilde g^{(7)}
- 2\rho \Lambda^3 \tilde g^{(6)}
- 2\rho \Lambda^3 \tilde g^{(5)} \tilde g^{(2)}
- 2\rho \Lambda^3 \tilde g^{(4)}  \tilde g^{(3)}
\nonumber \\ &&  \mbox{ }
+ 3 \rho^2 \Lambda^6 \tilde g^{(5)}
+ 6\rho^2 \Lambda^6  \tilde g^{(4)} \tilde g^{(2)}
+ 3\rho^2 \Lambda^6  \big( \tilde g^{(3)} \big)^2
 \nonumber \\ &&  \mbox{ }
+ 3\rho^2 \Lambda^6  \tilde g^{(3)}
\big(\tilde g^{(2)} \big)^2
- 4 \rho^3 \Lambda^9   \tilde g^{(4)}
- 12 \rho^3 \Lambda^9  \tilde g^{(3)} \tilde g^{(2)}
 \nonumber \\ &&  \mbox{ }
- 4 \rho^3 \Lambda^9 \big(  \tilde g^{(2)} \big)^3
+ 5 \rho^4 \Lambda^{12}  \tilde g^{(3)}
+ 10 \rho^4 \Lambda^{12} \big(  \tilde g^{(2)} \big)^2
 \nonumber \\ &&  \mbox{ }
- 6 \rho^5 \Lambda^{15}  \tilde g^{(2)}
+\rho^6 \Lambda^{18}.
\end{eqnarray}
A useful check on these is that the magnitude of the coefficients
must add to $2^{l-1}$.
These results for $l \ge 4$ rely upon the inferred form
for the corrected average.

\end{document}